\documentclass[a4paper, 10pt, twocolumn, superscriptaddress, aps, floatfix, accepted=2024-07-16]{quantumarticle}
\pdfoutput=1
\usepackage{amsmath, amssymb, graphicx, hyperref, xcolor, microtype, braket, mathtools}
\usepackage[export]{adjustbox}
\usepackage[numbers]{natbib}

\newcommand{\tmop}[1]{\ensuremath{\operatorname{#1}}}
\newcommand{\CX}{\textit{CX}}

\newcommand{\ZZ}{\textit{ZZ}}

\newcommand{\changes}[1]{#1}

\begin{document}

\title{Stabilization of symmetry-protected long-range entanglement in stochastic quantum circuits}

\author{Iosifina Angelidi}
\affiliation{Department of Physics and Astronomy, University College London, Gower Street, London, WC1E 6BT}

\author{Marcin Szyniszewski}
\affiliation{Department of Physics and Astronomy, University College London, Gower Street, London, WC1E 6BT}

\author{Arijeet Pal}
\affiliation{Department of Physics and Astronomy, University College London, Gower Street, London, WC1E 6BT}

\begin{abstract}
Long-range entangled states are vital for quantum information processing and quantum metrology. Preparing such states by combining measurements with unitary gates opened new possibilities for efficient protocols with finite-depth quantum circuits. The complexity of these algorithms is crucial for the resource requirements on a large-scale noisy quantum device, while their stability to perturbations decides the fate of their implementation. In this work, we consider stochastic quantum circuits in one and two dimensions comprising randomly applied unitary gates and local measurements. These operations preserve a class of discrete local symmetries, which are broken due to the stochasticity arising from timing and gate imperfections. In the absence of randomness, the protocol generates a symmetry-protected long-range entangled state in a finite-depth circuit. In the general case, by studying the time evolution under this hybrid circuit, we analyze the time to reach the target entangled state. We find two important time scales that we associate with the emergence of certain symmetry generators. The quantum trajectories embody the local symmetry with a time scaling logarithmically with system size, while global symmetries require exponentially long times. We devise error-mitigation protocols that significantly lower both time scales and investigate the stability of the algorithm to perturbations that naturally arise in experiments. We also generalize the protocol to realize toric code and Xu-Moore states in two dimensions, opening avenues for future studies of anyonic excitations. Our results unveil a fundamental relationship between symmetries and dynamics across a range of lattice geometries, which contributes to a broad understanding of the stability of preparation algorithms in terms of phase transitions. Our work paves the way for efficient error correction for quantum state preparation.
\end{abstract}

\maketitle

\section{Introduction}

Entanglement is a resource that plays an essential role in the exponential speedup of quantum algorithms~\cite{Jozsa1997, Ekert1998, Jozsa2003, Ding2007, Kendon2004, Aaronson2015, Huang2018}, the robustness of quantum teleportation~\cite{Yin2020}, and quantum error correction~\cite{Brun2006, Reed2013entanglement}. Essentially, all practical applications of entanglement demand shielding from external noise, for which quantum error correction is indispensable~\cite{Knill1997, Gottesman2001, Aaronson2015, Campbell2017}.  Some of the proposals for error-protected logical qubits require stable long-range and multipartite entanglement, such as surface codes~\cite{Kitaev2003, Dennis2002, Fowler2012}. These entangled states appear to be extremely hard to find and manipulate in nature. Therefore, preparing them artificially using current digital quantum simulators may allow the implementation of quantum error correction beyond break-even in the near term~\cite{Ofek2016, Arute2019, Guo2021, Sivak2023, Ni2023}. On the path to universal quantum computing, the preparation of these states can be an important milestone and may have an impact on quantum teleportation and cryptography~\cite{Yin2020}.

\begin{figure}[b]
  \centering
  \includegraphics[width = 0.99\columnwidth]{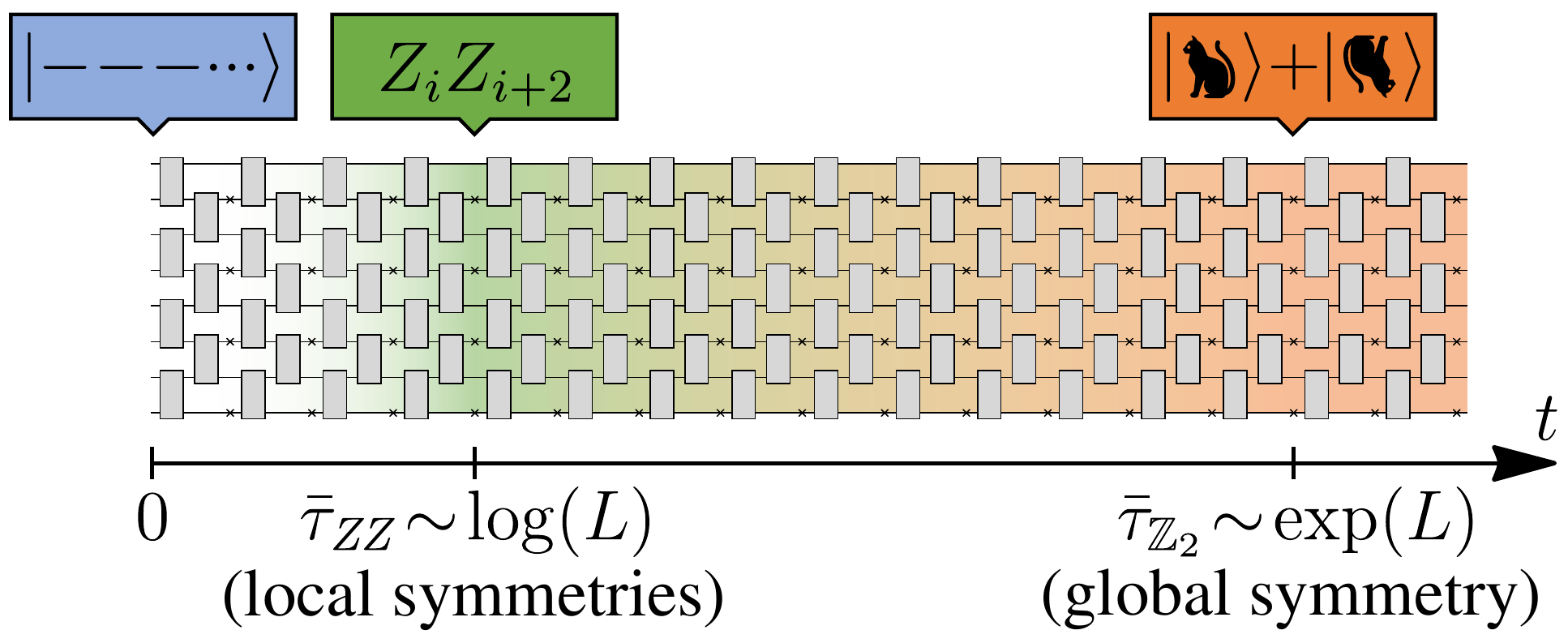}
  
  \caption{
  Diagram of the stochastic quantum circuit studied in this work, where the unitary evolution (gates in grey) is interspersed with measurements (crosses). The initial state is a product state in the $X$-basis. After time $\bar\tau_\ZZ$, which is logarithmic in the system size $L$, the state exhibits local $Z_i Z_{i+2}$ symmetries (in green). The state turns into a long-range entangled cat state (in orange) after time $\bar\tau_{\mathbb{Z}_2}$, which is exponential in $L$, when the global $\mathbb{Z}_2$ symmetry emerges.
  \label{fig:summary}}
\end{figure}

Preparing long-range entangled (LRE) states in experiments using quantum gates and measurements is a challenging task, since it is directly linked with the control of a large number of qubits and maintaining their quantum correlations~\cite{Cory1998, Aaronson2015}. Therefore, developing efficient state preparation protocols to engineer LRE states has attracted significant attention recently. Most of the efforts were focused on realizing such states as ground states of quantum many-body Hamiltonians, where long-range entanglement is protected by the symmetries in the Hamiltonian. The complexity and resource requirement for preparing LRE states through unitary time-evolution via a Hamiltonian or a quantum circuit renders their implementation inefficient. The circuit depth typically grows extensively with the system size for local operations~\cite{Bravyi2006, Aguado2008, Konig2009, Chen2010, Zaletel2020, Soejima2020, Liu2022, Wei2022} and can be reduced to log-depth for nonlocal operations~\cite{Cruz2019, Liao2021, Konig2009, Aharonov2018}. However, there is a fundamental limit to these preparation algorithms due to the Lieb-Robinson bound which restricts state preparation protocols using unitary circuits~\cite{Bravyi2006, Hastings2010}.

Interestingly, the breaking of unitarity through quantum measurements allows access to new dynamical phases of matter and phase transitions~\cite{Verstraete2009, Diehl2011, Sang2021}.  This opens a new regime for investigating quantum many-body systems in state-of-the-art quantum simulators. In the presence of symmetries, introducing measurements in a unitary circuit for LRE state preparation can reduce the circuit depth substantially, which allows efficient implementation in large systems~\cite{Briegel2001, Verresen2021, Raussendorf2005, Brennen2008, Lanyon2013, Bolt2016, Piroli2021, Li2021decoder, Lu2022}. Measurement outcomes interspersed through the circuit can feed forward into the hybrid evolution that is closely connected to quantum error-correcting codes~\cite{Gottesman2001, Dennis2002}. Remarkably, time-evolving an initially separable state via a sequence of local two-site unitary entangling gates, gives rise to a short-range entangled state, for example, the so-called cluster state~\cite{Briegel2001}. Subsequently, applying single-site measurements on a subsystem leaves the unmeasured degrees of freedom in a symmetry-protected long-range entangled state with high fidelity. For instance, the realization of the GHZ state and the toric code state is possible, by performing measurements on the 1D and 2D cluster state respectively~\cite{Briegel2001, Verresen2021}.

However, the stability of the protocols to errors in the application of gates and measurements is a significant factor for efficient, high-fidelity state preparation. 
Quantum fault-tolerance thresholds have been recently probed in hybrid random circuits~\cite{Li2018, Skinner2019, Chan2019, Szyniszewski2019, LiVasseur2021, Shtanko2020}, which allow direct access to the entanglement structure. 
In this work, we are considering the effects of imperfections in the application of unitaries and measurements and hence turning the state preparation protocol into a stochastic process.  This takes into consideration different kinds of errors that can naturally arise in a state preparation experiment, and which preserve but also break the symmetries. Through the study of the dynamics of the stochastic circuit under the influence of errors, our analysis characterizes the deviations in run time to converge to the LRE state and their interplay with the local and global symmetries, and helps identify the quantum fault-tolerance thresholds of the model. It is non-trivial that the target state is achieved for all quantum trajectories independent of the circuit realization. 

In the presence of errors, the hybrid dynamics leads to excursions away from the LRE states. We find two characteristic time scales for the emergence of local and global symmetries in the late time state. A time scale \textit{logarithmic} in system size is associated with the evolving state exhibiting the local stabilizer symmetry. While in order to realize the global symmetric state the system takes an \textit{exponentially} long time (see Fig.~\ref{fig:summary} for the intuitive picture). The exponential time scale is cut off if either unitaries or measurements are perfectly applied with no errors. An analytical understanding of these time scales is developed through a weighted graph connecting the errors. We consider the stability of the target state under perturbations to the unitary elements of the circuit and find a measurement-induced phase transition between area-law and volume-law entangled states. We generalize our results to two-dimensional lattices, where we find states that can serve as topological quantum error correction codes.

This work is structured as follows.
First, in Sec.~\ref{sec:timescales}, we present a LRE state preparation protocol with measurements, that achieves a cat state with high fidelity in a finite-depth circuit. We then consider the effect of randomly applied gates and measurements, and provide an in-depth explanation of our analytical and numerical results. 
In Sec.~\ref{sec:speedup}, we explore three different methods to speed up our stochastic protocol. These methods include the application of a local decoder, a protocol where we halt the application of unitaries when a specific fidelity is achieved, and lastly by enforcing the global symmetry. Sec.~\ref{sec:stability} is focused on the stability of the protocol under various imperfections: timing imperfections in the unitary evolution, and an additional interaction term that leads to a measurement-induced phase transition. In Sec.~\ref{sec:2d}, we broaden our scope further and study a similar stochastic protocol in two-dimensional lattices, which support topological error-correcting states. We discuss our main findings in Sec.~\ref{sec:discussion} and talk about future work.

\section{Long-range entanglement in stochastic circuits}
\label{sec:timescales}

\subsection{Exact case}

The preparation of LRE states in local unitary circuits demands extensive depths~\cite{Bravyi2006}, unless one inserts measurements in such protocols, which overcomes this barrier and leads to a finite-depth circuit. Here, we discuss an example of such a state preparation protocol that was first introduced in Ref.~\cite{Briegel2001}, which uses a finite depth circuit with measurements in order to obtain a cat state with high fidelity. This protocol was generalized recently to states with non-abelian topological order~\cite{Verresen2021}.

\begin{figure}[tb]
  \centering
  \includegraphics[width = 0.99\columnwidth]{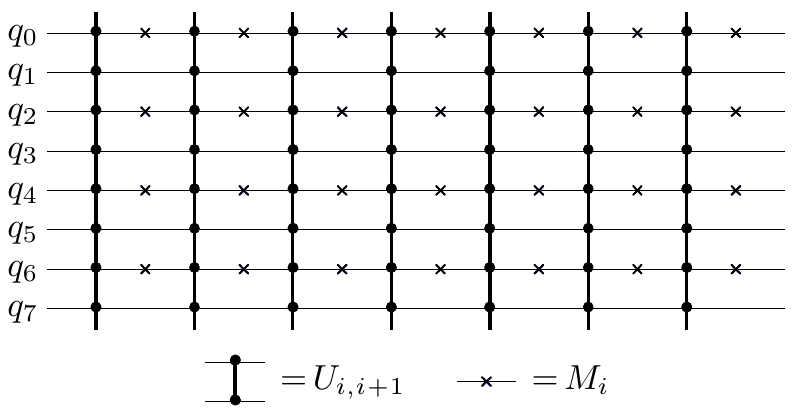}
  
  \caption{Quantum circuit (time on the $x$ axis, space on the $y$ axis) showing evolution using unitary evolution (vertical lines ending in dots) and measurements (crosses).\label{fig:circuit-exact}}
\end{figure}

Let us consider a 1D stabilizer circuit of length $L$, where we start from the all-minus state $| \psi \rangle = |{-} {-} {-} \,{\cdots} \rangle = | - \rangle^{\otimes L}$, where $|-\rangle$ is the eigenstate of Pauli $X$ operator with an eigenvalue of $-1$. We evolve the system using the unitary evolution
\begin{align}
  U &= \exp \left( - i \Delta t H \right) \nonumber \\
  &= \prod_i \exp \left( - i \frac{\pi}{4} Z_i Z_{i + 1} \right) = \prod_i U_{i, i + 1}, \label{eq:unitary}
\end{align}
where the Hamiltonian is $H = \sum_i Z_i Z_{i + 1}$ and the evolution time is set to $\Delta t=\pi/4$.
This is followed by measurements in the $X$ direction applied only on even sites, $M = \prod_{i\text{ even}} M^X_i$. The unitary evolution alone produces a cluster state when using $\Delta t = \pi/4$~\cite{Briegel2001, Verresen2021} -- a state with stabilizers of the form $X_i \prod_{j\in E(i)} Z_j$, where $E(i)$ designates the immediate neighborhood of the site $i$. In the case of 1D circuit, this leads to the ${ZXZ}$ symmetry,
\begin{equation}
  U | \psi \rangle = Z_{i - 1} X_i Z_{i + 1} U | \psi \rangle.
\end{equation}
Followed by a single layer of measurements, results in a cat state on odd sites, while even sites become separable. The stabilizers of the cat state are $\pm Z_{i}Z_{i+2}$ \changes{($i$ is odd)} and a global symmetry of $\prod_{i\text{ odd}} X_i$ (parity of spins in the $X$-direction on odd sites). Subsequent applications of this protocol (see Fig.~\ref{fig:circuit-exact}) do not disturb the cat state. \changes{Note that starting from the product state in the $X$ basis is crucial. One could be tempted to start instead from the $Z$-basis product state, where the $Z_i Z_{i+2}$ stabilizers are already present -- however, this state lacks the global symmetry, and its evolution does not produce entanglement.}

Interestingly, the outcomes of the measurements can be used to harmonize the signs of the stabilizers, and to produce a GHZ state~\cite{Briegel2001, Verresen2021}. In this work, however, we will omit the discussion of the signs, as these are not important to the entanglement structure of the state. It is worth noting, that the cluster state consists is an example of a symmetry-protected topological (SPT) phase which could be measured to achieve a LRE state~\cite{Tantivasadakarn2021}.

\subsection{Stochastic case}

Here we discuss the problem of achieving long-range entangled cat states in circuits where unitary evolution and measurements are applied with probabilities $p_u$ and $p_m$, respectively. \changes{In this work, ``errors'' are defined as the absence of a gate or a measurement, which can be thought of as failed gates and unsuccessful measurements.} This is a simple model of errors that are potentially relevant for experiments, where the failed measurements are assumed to give a random output. \changes{We assume no knowledge of the locations of errors, unless specified otherwise.} The main purpose of introducing stochasticity is to test the thresholds for errors in gates and measurements. Similar models have been studied within the context of quantum error correction, where random unitary circuits with intermittent measurements have been used to characterize entanglement and purification phase transitions, which can provide insights into the threshold for quantum error correction~\cite{Choi2020, Fan2021, Li2021, Li2021decoder}. More specifically, the transition is driven by the error rate, i.e.\@ random measurements, in these circuits, where at a finite rate of measurement the system undergoes a measurement-induced phase transition~\cite{Li2018, Skinner2019, Chan2019, Szyniszewski2019, LiVasseur2021}.

We first consider the gates from within the Clifford group, while the measurements are Pauli operators. This preserves the stabilizer states, and can be implemented efficiently on a classical computer using the Aaronson-Gottesman algorithm~\cite{Aaronson2004}, which employs the tableau formalism to represent the states. We convert all Clifford gates to a series of Hadamard gates $H$, phase gates $S$, and CNOT gates $\CX$ (generators of the Clifford group) using the algorithm presented in Ref.~\cite{Niemann2014}. The gates used in this work include:
\begin{align}
  \exp \left( - i \frac{\pi}{4} Z_a Z_b \right) & = \frac{1}{\sqrt{2}} (1 -
  i Z_a Z_b) \\
  & = \frac{1-i}{\sqrt 2} H_a \CX_{b a} H_a S_a S_b, \nonumber \\
  \exp \left( - i \frac{\pi}{4} X_a X_b \right) & = 
  \frac{1-i}{\sqrt 2} H_a H_b S_b H_b \CX_{ab} S_a H_a. 
\end{align}
Measurements of any Pauli string can be implemented by deconstructing the corresponding projector into gates applied on the projectors of $Z$ measurement. The measurement of $X$ can be implemented with the addition of Hadamard gates,
\begin{equation}
  M^X_a = H_a M^Z_a H_a.
\end{equation}

\begin{figure}[tb]
  \centering
  \includegraphics[width=0.99\columnwidth]{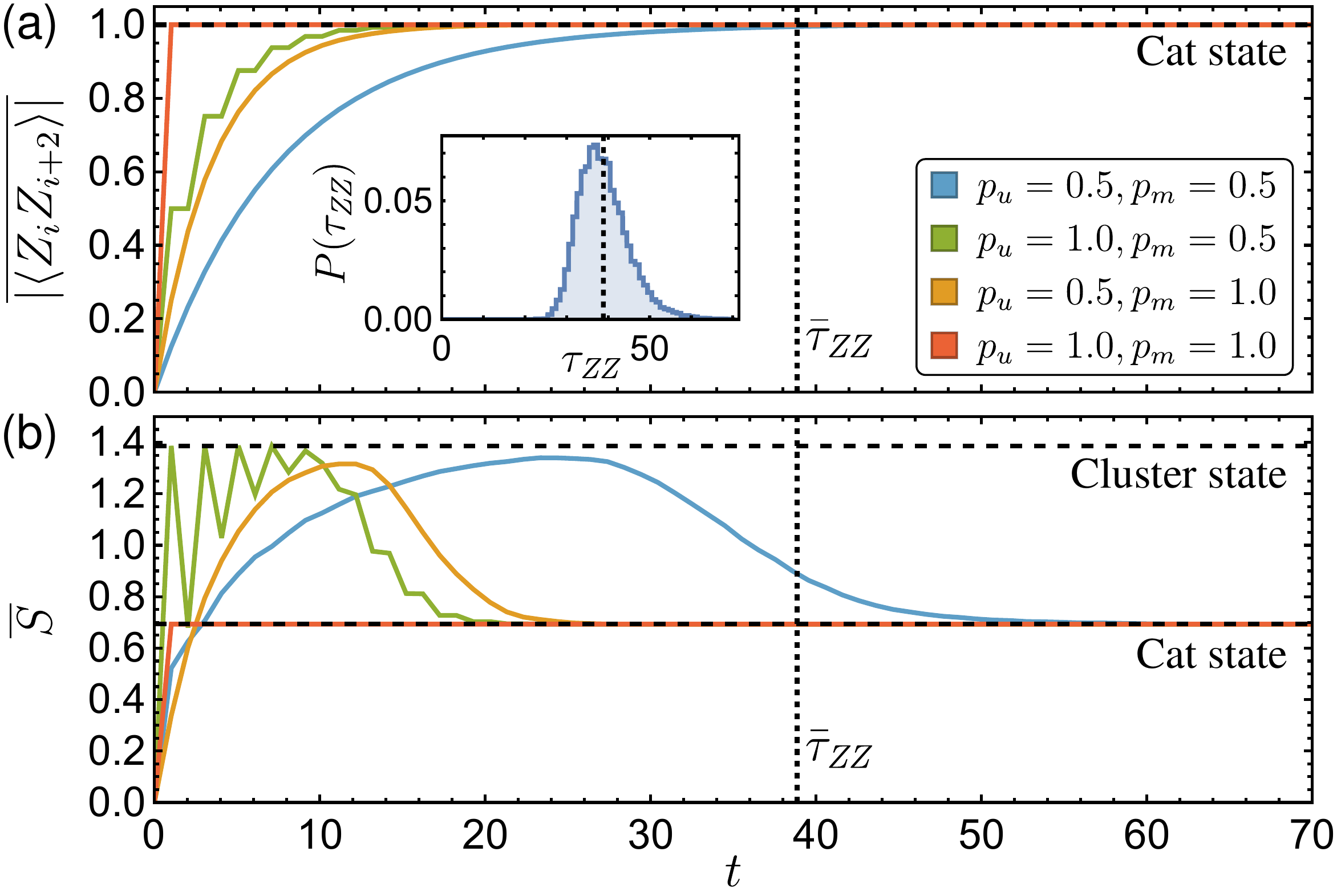}
  \caption{Initial behavior of the stochastic circuit: (a)~the expectation value $|\langle Z_i Z_{i+2} \rangle|$ and (b)~half-chain von Neumann entanglement entropy $S$, averaged over 1000 realizations for the system size $L=512$. Dashed lines show expected values for the Schr\"odinger cat state and the cluster state. Dotted lines show the mean time $\bar\tau_\ZZ$ needed for the state to exhibit $Z_i Z_{i+2}$ stabilizers, for $p_u = 0.5$ and $p_m = 0.5$. The inset in~(a) shows the distribution of time $\tau_\ZZ$ for the same parameters.
  \label{fig:initial_time}}
\end{figure}

Using this computational toolkit, we now consider the behavior of the circuit with errors in the application of gates and measurements. More specifically, we apply the gates with probability $p_u$ and measurements with probability $p_m$, where both can be tuned below 1. \changes{The measurements only act on even sites.} First, we study the expectation value of $|\langle Z_i Z_{i+2} \rangle|$ \changes{(where $i$ is odd)}, which gives us information about the existence of $\ZZ$ stabilizers, i.e.\@ the local stabilizers of the cat state. As seen in Fig.~\ref{fig:initial_time}(a), these stabilizers emerge at long times, where the timescales depend nontrivially on both $p_m$ and $p_u$.

We also discuss the behavior of the von Neumann entanglement entropy between subsystem A and its complement B, defined as $S =\nobreak - \rho_\text{A} \log \rho_\text{A}$, where $\rho_\text{A} = \tmop{tr}_B \rho$ is the reduced density matrix of subsystem A calculated by tracing out the complement B from the density matrix $\rho = |\psi\rangle\langle\psi|$. For stabilizer circuits, the entropy can be calculated by taking the stabilizer rows of the tableau and removing columns and rows corresponding to complement B, as well as the sign column, leaving matrix $\mathcal{M}$. Then, $ S = (\tmop{rank}(\mathcal{M}) - L_A) \log 2$, where $L_A$ is the length of subsystem A, and the rank is taken over binary numbers (i.e.\@ rank over mod 2)~\cite{Hamma2004, Hamma2005, Nahum2017}.

Measuring the mean of the half-chain entanglement entropy $S$ (where the subsystem is half of the entire chain), we observe that generically the state may start in an almost separable state of $S \sim 0$, then the entanglement grows close to a cluster state value with $S \sim 2 \log (2)$, and finally, at long times the state exhibits entanglement structure similar to a cat state with $S \sim \log (2)$ [see Fig.~\ref{fig:initial_time}(b)]. We note that the long-time steady state of this circuit is a low-entangled state with $S \sim \log (2)$, satisfying the \textit{area law}. An interesting situation happens when $p_m = 0$ and $p_u = 1$, when the state oscillates between a cluster state $(S = 2 \log 2)$ and a separable all-minus state $(S = 0)$. The temporal behavior of the stochastic circuit suggests that at long times a cat state is produced. 

We now aim to estimate the mean time to achieve a Schr\"odinger cat, or equivalently, an average minimum depth of the circuit. The distribution of this depth is directly related to the notion of quantum complexity~\cite{Araujo2021, Rindell2023, Suzuki2023} -- in state preparation protocols, complexity is usually defined through the minimum depth required to represent a target state. Hence, by characterizing the time distribution we reveal information about the quantum complexity of this problem.

\begin{figure}[tb]
  \centering
  \includegraphics[width=0.99\columnwidth]{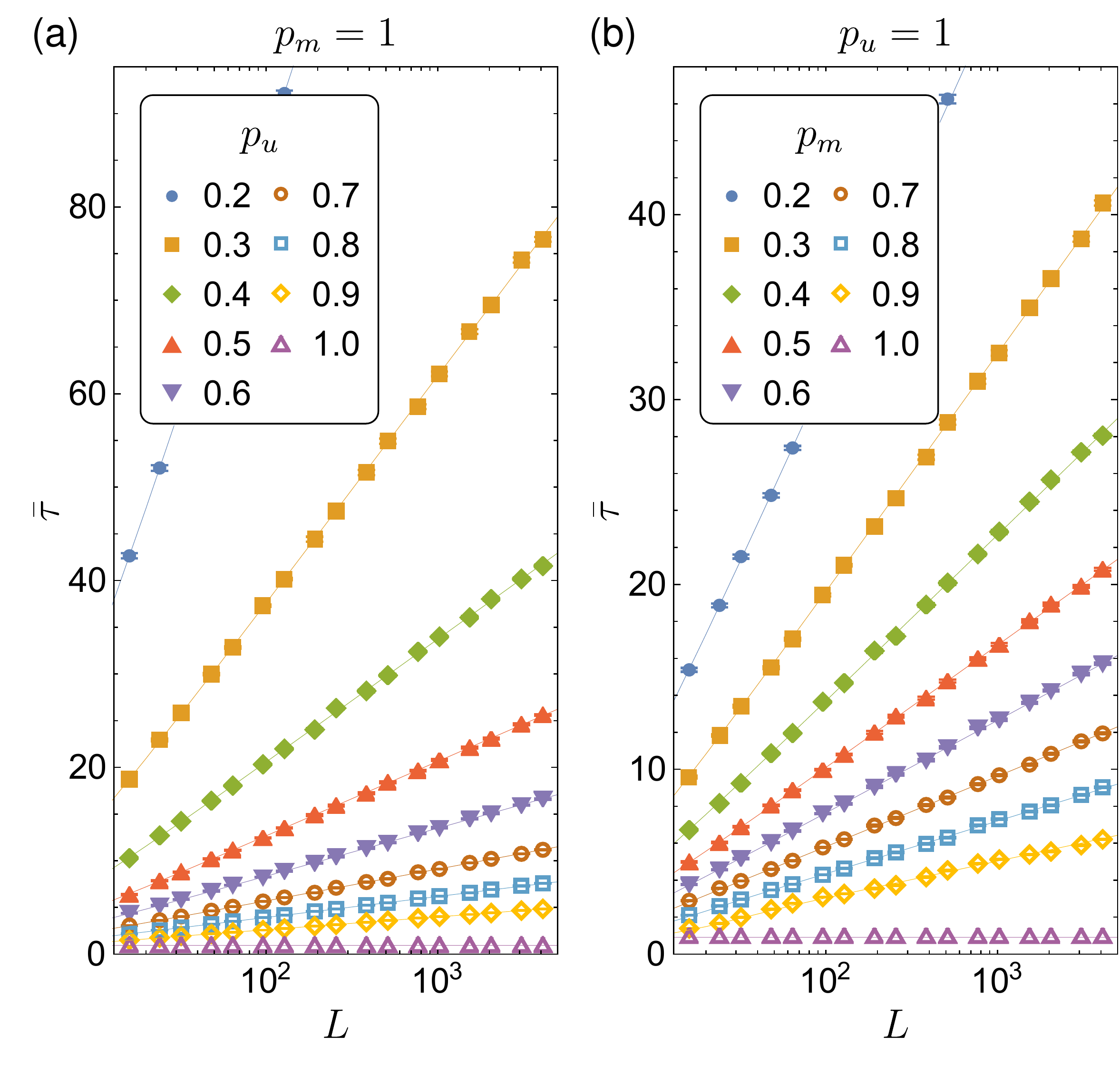}
  \caption{Mean time $\bar\tau$ of achieving a cat state in a stochastic circuit with  unitaries of frequency $p_u$ and measurements of frequency $p_m$. When either (a)~$p_m = 1$ or (b)~$p_u = 1$, we find that $\bar\tau \sim \log(L)$. Lines are analytical estimates from Eqs.~(\ref{eq:exact-mean-time-pm1}) and (\ref{eq:exact-mean-time-pu1}).
  \label{fig:mean_time_pupm1}}
\end{figure}

The upper bound on the mean time $\bar \tau$ is a naive situation when in a certain circuit layer all possible unitaries and measurements are applied, i.e.\@
\begin{equation}
  \bar \tau \lesssim 1 / [(p_u)^L (p_m)^L] = (p_u p_m)^{-L} \sim \exp (L),
\end{equation}
which scales \textit{exponentially} with the system size. Despite this, we find that for two scenarios: $p_u = 1$ or $p_m = 1$, the mean time to achieve a cat state is \textit{logarithmic} with the system size, $\bar\tau \sim \log(L)$, see Fig.~\ref{fig:mean_time_pupm1}. Only when both $p_u < 1$ and $p_m < 1$, our results [see Fig.~\ref{fig:mean_time_pupmbelow}] show exponential growth for large system sizes, $\bar\tau \sim \exp(L)$, as the upper bound would suggest.

To understand the origin of the logarithmic and exponential time scales present in this stochastic state preparation protocol, we proceed by investigating in detail the dynamics of the circuit for each scenario.

\subsubsection{Logarithmic divergence for \texorpdfstring{$p_m = 1$}{pm = 1}}

\begin{figure}[tb]
  \centering
  \includegraphics[width=0.99\columnwidth]{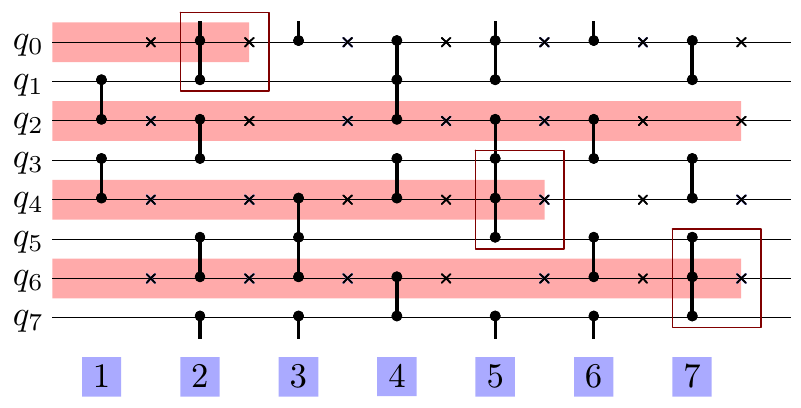}
  
  \caption{Example of a circuit with $p_m = 1$ and $p_u < 1$. Red areas show   where the $Z_{i} Z_{i + 2}$ symmetries are absent, and blue numbers designate the layers. Any two unitaries neighboring the measurement are responsible for fixing a $Z_{i} Z_{i + 2}$ symmetry (highlighted by red outlines). Note that one needs to fix $(L / 2 - 1)$ stabilizers, as the last stabilizer follows naturally from combining other $Z_{i} Z_{i + 2}$ stabilizers. \label{fig:circuit-pm1}}
\end{figure}

The behavior of the mean time $\bar \tau$ can be probed through the emergence of cat state stabilizers: local $Z_{i}Z_{i+2}$ \changes{($i$ odd)} and a global symmetry of $\prod_{i\text{ odd}} X_i$. In the case of $p_m = 1$, we can first note that the $\mathbb{Z}_2$ symmetry of $\prod_{i\text{ odd}} X_i$ is always preserved during this evolution. Furthermore, a $Z_i Z_{i+2}$ stabilizer can be produced when two unitaries neighboring a measurement are applied (cf.\@ red outlines in the diagram in Fig.~\ref{fig:circuit-pm1}). This stabilizer is then stable towards further evolution and measurements. One can note that to produce the cat state, we only need to fix $(L/2-1)$ of $Z_i Z_{i+2}$ stabilizers, as the last stabilizer follows by multiplying all other $Z_i Z_{i+2}$ stabilizers.

A rough approximation of the mean time to produce all the cat state stabilizers, $\bar\tau$, can be performed using the following argument. After a single layer, the number of sites that did not have two consecutive unitaries applied to them is $L / 2 \times (1 - p_u^2)$, on average. When the circuit reaches depth $t$, this number is $L / 2 \times (1 - p_u^2)^t$. If this number reaches unity, then we should reach the cat state. This leads to
\begin{equation}
  \bar\tau \sim \frac{\log (2 / L)}{\log (1 - p_u^2)}.
  \label{eq:naive-mean-time}
\end{equation}
A more precise extraction of $\bar\tau$ can be done by treating each application of neighboring unitaries separately as a random variable. This calculation can be found in detail in Appendix~\ref{sec:appendix-log-time}, which yields additional correction terms,
\begin{equation}
  \bar\tau \approx \frac{\log (2 / L)}{\log (1 - p_u^2)} - \frac{\gamma - 1}{\log (1 - p_u^2)} + \frac{1}{2},
  \label{eq:log-mean-time}
\end{equation}
where $\gamma$ is the Euler-Mascheroni constant. The more precise expression from Eq.~(\ref{eq:exact-mean-time-pm1}) is plotted in Fig.~\ref{fig:mean_time_pupm1}(a) as solid lines and it agrees fully with our numerical calculations.

\begin{figure}[tb]
  \centering
  \includegraphics[width = 0.99\columnwidth]{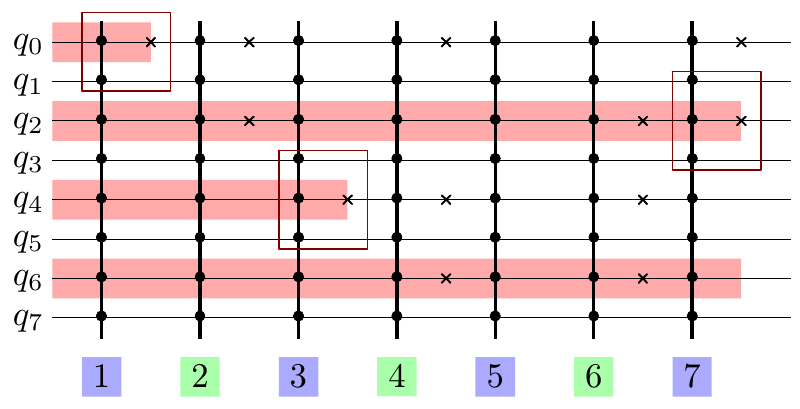}
  
  \caption{Example of a circuit with $p_u = 1$ and $p_m < 1$. Red areas show the absence of the $Z_{i} Z_{i + 2}$ symmetry. We find that measurements in odd layers (in blue) are responsible for fixing the $Z_{i} Z_{i + 2}$ symmetry, while the measurements in even layers (in green) are unimportant.\label{fig:circuit-pu1}}
\end{figure}

Using the cumulative distribution function derived in this calculation, we can also write the time needed to achieve a certain fidelity $\phi$ (an average percentage of states that will be Schr\"odinger cat states),
\begin{equation}
  \tau(\phi, p_u^2) \sim \frac{1}{2}  \frac{\log \left( \frac{8 (1 - \phi)}{L (L - 2)} \right)}{\log (1 - p_u^2)} - 1.
  \label{eq:time-fidelity}
\end{equation}
Importantly, this time is also logarithmic in the system size, which provides a controlled protocol for achieving a high-fidelity cat state in a simulation experiment.

\subsubsection{Logarithmic divergence for \texorpdfstring{$p_u = 1$}{pu = 1}}
\label{sec:cat-pu1}

The case of $p_u = 1$ is similar to that of $p_m=1$, with a few exceptions. The role of applying two unitaries is now taken by an application of measurement, i.e.\@ $p_u^2$ is replaced by $p_m$. Additionally, if a measurement is applied on $i$-th site in an odd layer, this produces a $Z_{i - 1} Z_{i + 1}$ stabilizer, which then is stable to further evolution and measurements. However, a measurement is not producing a stable stabilizer if it is applied in an even layer (see the diagram in Fig.~\ref{fig:circuit-pu1}). Accounting for this, the mean time $\bar\tau$ is given by:
\begin{equation}
  \bar\tau \approx  2 \left( \frac{\log (2 / L)}{\log (1 - p_m)} - \frac{\gamma - 1}{\log (1 - p_m)} \right),
\end{equation}
which is again logarithmic in the system size. Similarly, the time needed to achieve a cat state given certain fidelity can be estimated as $2\tau(\phi, p_m)$ from Eq.~(\ref{eq:time-fidelity}), where the factor of 2 comes from the even-odd layer effects.

\subsubsection{Fast divergence for \texorpdfstring{$p_u < 1$}{pu < 1} and \texorpdfstring{$p_m < 1$}{pm < 1}}

\begin{figure}[tb]
  \centering
  \includegraphics[width = 0.99\columnwidth]{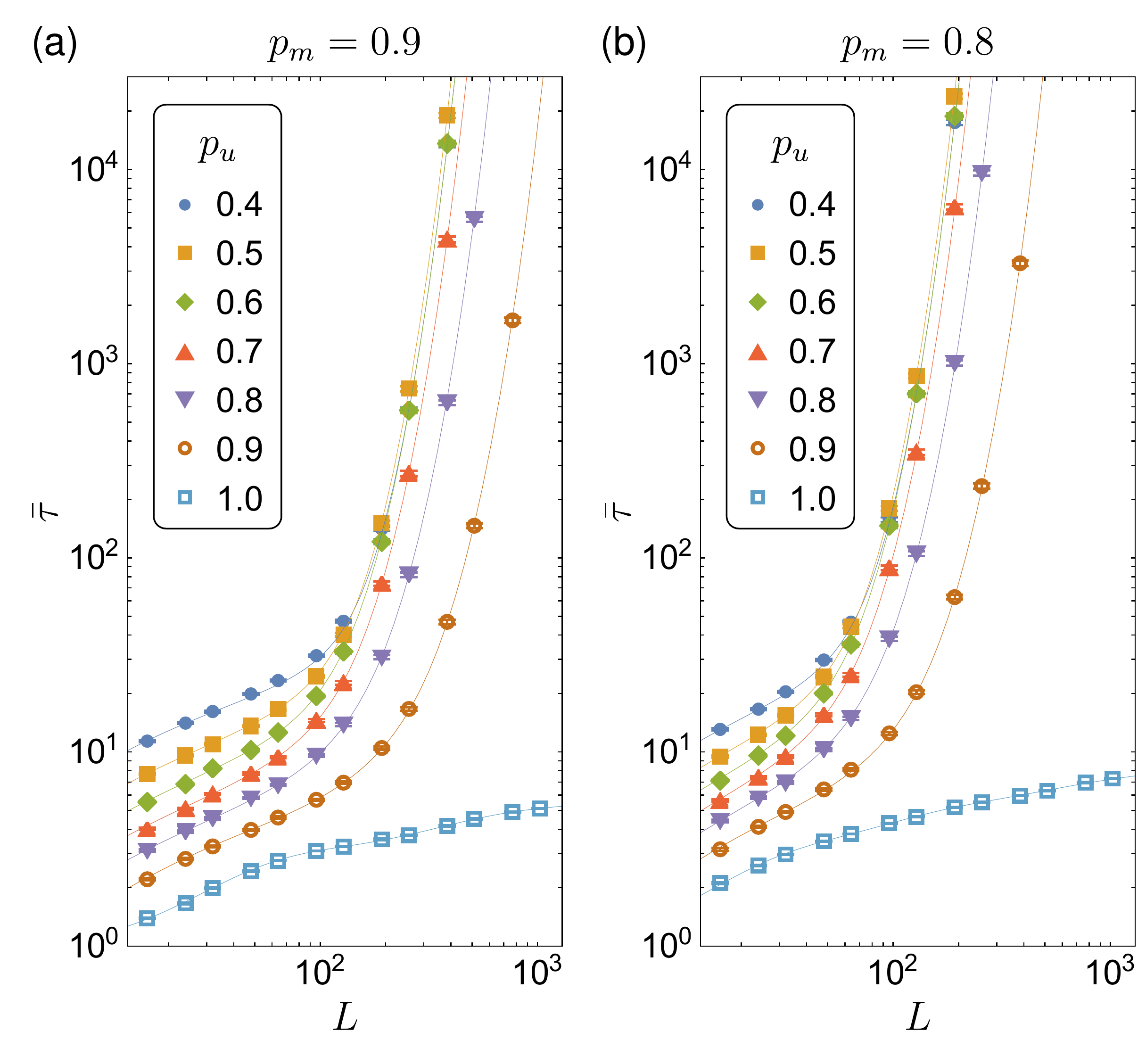}
  \caption{Mean time of achieving a cat state in a stochastic circuit with unitaries of frequency $p_u$ and measurements of frequency $p_m$. When $p_u<1$ and $p_m<1$, we find exponential divergence of $\bar\tau$ for large system sizes. Solid lines are analytical predictions based on Eqs.~(\ref{eq:CDF_ZZ}) and (\ref{eq:mean-time-Z2}).
  \label{fig:mean_time_pupmbelow}}
\end{figure}

We now turn to the discussion of both $p_u<1$ and $p_m<1$. First to note is that the evolution does not preserve the $\mathbb{Z}_2$ symmetry on the odd sites. The individual $Z_i Z_{i + 2}$ stabilizers \changes{(where $i$ is odd)} get locked similarly to the cases of $p_m = 1$ or $p_u = 1$, so that it takes a logarithmic time to reach $Z_i Z_{i+2}$ stabilizers on every pair of odd sites. However, one needs to wait an exponentially long time to recover the $\mathbb{Z}_2$ symmetry of $\prod_{i \text{ odd}} X_i$. This generically results in an exponential mean time $\bar\tau$ for larger circuits, as seen in Fig.~\ref{fig:mean_time_pupmbelow}.

\begin{figure}[tb]
  \centering
  \includegraphics[width = \columnwidth]{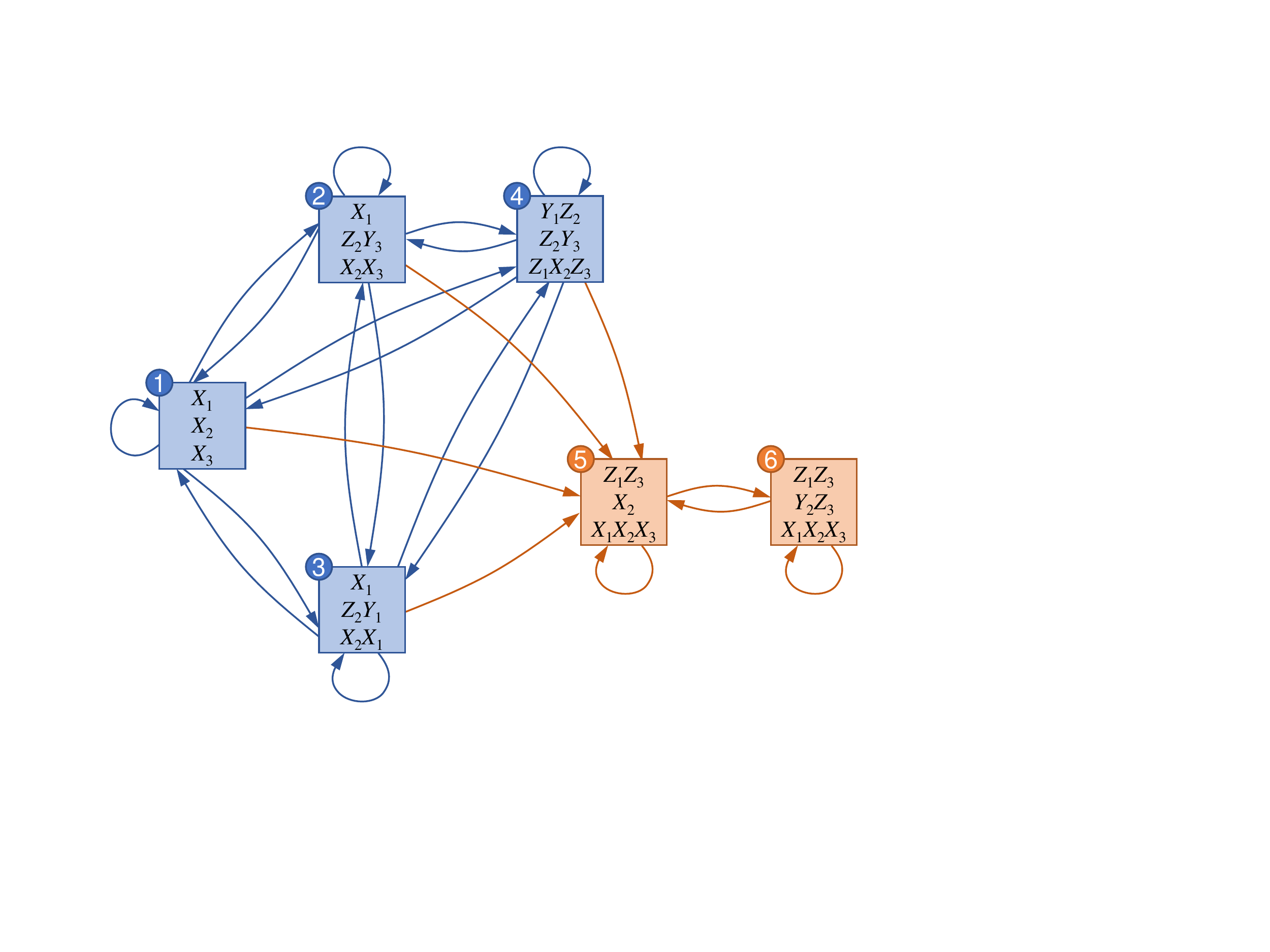}
  
  \caption{Weighted graph showing six possible 3-qubit states (each state is defined through stabilizer generators). Edges correspond to possible transitions between different states. Vertices 5 and 6 (in orange) are stable at long times, and both correspond to states with $Z_i Z_{i+2}$ stabilizers. \label{fig:transfer-graph}}
\end{figure}

A more in-depth analysis can be performed by noticing that the chain can be split into 3-site clusters. For odd $i$, both unitaries and measurements supported on sites $(i, i+1, i+2)$ commute with the unitaries and measurements on sites $(i+2, i+3, i+4)$. Therefore, except for one global stabilizer that is always preserved ($\prod_{i = 0}^{L - 1} X_i$), one can always write other independent stabilizers as 3-site operators. Thus, we can consider every possible transition between all 3-site clusters in the circuit, which can be visualized as a weighted graph in Fig.~\ref{fig:transfer-graph}, where each state is represented by its stabilizer generators. For example, the transition from Vertex 1 to itself (the $X$-basis state remains an $X$-basis state) can happen in two scenarios: when only one unitary is applied and measurement is performed, and when no unitaries are applied (and measurement is either performed or not). The transition probability for this process is, therefore, $[2 p_m p_u (1-p_u)+(1-p_u)^2]$. This idea of exploring different probabilities of transitions from one state to another is similar to noise channels in quantum error correction~\cite{Lidar2013, Nielsen2010}.

The weighted graph provides an intuitive picture of the origin of both the logarithmic and the exponential scalings. The initial state (all-minus state, Vertex 1) always transitions into any of the two orange vertices at long times, achieving the $Z_i Z_{i+2}$ stabilizer with a certain nonzero probability. This directly leads to the logarithmic mean time to achieve $Z_i Z_{i+2}$ stabilizers in the entire circuit,
\begin{equation}
  \bar\tau_{\ZZ} \sim \log(L).
  \label{eq:mean-time-zz}
\end{equation}
Secondly, to achieve the global $\mathbb{Z}_2$ symmetry of $\prod_{i\text{ odd}} X_i$, one needs to combine the global $\prod_{i = 0}^{L - 1} X_i$ symmetry with local $X_i$ stabilizers, which can only be achieved when every possible 3-site state is Vertex 5. The graph can be used to estimate a probability $p_X$ of obtaining one local $X_i$ stabilizer,
\begin{equation}
  p_X = \frac{p_m + 2 (1 - p_m) p_u (1 - p_u)}{p_m + 4(1 - p_m)p_u(1 - p_u)},
\end{equation}
which then leads to the time $\bar\tau_{\mathbb{Z}_2}$ of achieving a $\mathbb{Z}_2$ symmetry of $\prod_{i\text{ odd}} X_i$,
\begin{equation}
  \bar\tau_{\mathbb{Z}_2} \sim p_X^{- L / 2} \sim \exp (a L).
  \label{eq:mean-time-Z2}
\end{equation}
See Appendix~\ref{sec:appendix-exp-time} for detailed derivations.

To summarize, when $p_u<1$ and $p_m<1$ the mean time $\bar\tau$ needed for the cat state consists of two timescales (see Fig.~\ref{fig:summary}): (1)~the time needed for the local $Z_i Z_{i+2}$ stabilizers, which scales logarithmically with system size, $\bar\tau_\ZZ \sim \log(L)$, and (2)~the time needed to recover the global $\mathbb{Z}_2$ symmetry of $\prod_{i \text{ odd}} X_i$, which scales exponentially with the system size, $\bar\tau_{\mathbb{Z}_2} \sim \exp(L)$. This implies that away from the exact limits of $p_u = 1$ or $p_m=1$, the system enters a different manifold of state preparation time scales and quantum complexity. We compare numerical results with these analytical insights in Fig.~\ref{fig:mean_time_pupmbelow}) and we find that our analytical predictions are consistent with the numerics (cf.\@ solid lines vs markers).

\changes{Interestingly, assuming that we know the locations of measurements, one can synchronize the signs of the stabilizers of the cat state, turning it into a coherent GHZ state. To do this, we note the last measurement outcome for each even site, and then flip the spins using $X$ gates analogous to the protocol of Ref.~\cite{Verresen2021}.}

The emergence of the two timescales is of practical importance, as it implies that certain local properties of the long-range entangled states (like the $Z_i Z_{i+2}$ symmetries) can be achieved with an efficient protocol in the presence of noise. \changes{Generation of stabilizers with support larger than a single site is important for the presence of long-range entanglement.} The circuit depth needed \changes{for this} can be easily estimated using our simple formula for $\tau (\phi, p_u^2 p_m)$ from Eq.~(\ref{eq:time-fidelity}). On the other hand, the existence of exponential scaling can hinder the ability to produce desired states in experimental setups, especially states endowed with certain global symmetries. Our results guide us to consider different speed improvements of this protocol by tackling either of the time scales.

\section{Error mitigation protocols}
\label{sec:speedup}

In this section, we propose protocols for mitigating the disruptive effects of errors and lowering the mean time needed to achieve a long-range entangled Schr\"odinger cat state. We can influence both the $\log(L)$ growth, which dominates for small systems and is related to local stabilizers, or the $\exp(L)$ growth, which dominates at large systems and signifies the recovery of a global symmetry.

\subsection{Local decoding of measurement outcomes}

Quantum decoders are key ingredients in quantum error correction when it comes to identifying and correcting errors~\cite{Li2021decoder, Delfosse2021, Herold2017, Cong2022, Chamberland2022, Delfosse2021quantum}. They are used in various quantum computing applications including quantum cryptography and communication where fault tolerance and accuracy are crucial. An important ingredient of a decoder is the ability to adapt given the current state of the circuit (e.g.\@ a measurement outcome), effectively conditioning the future gates on the current quantum information. Usually, local decoders are used due to their relatively easy experimental implementation, however, efficient global decoders can be also implemented~\cite{Chamberland2022}.

\begin{figure}[tb]
  \centering
  \includegraphics[width = 0.99\columnwidth]{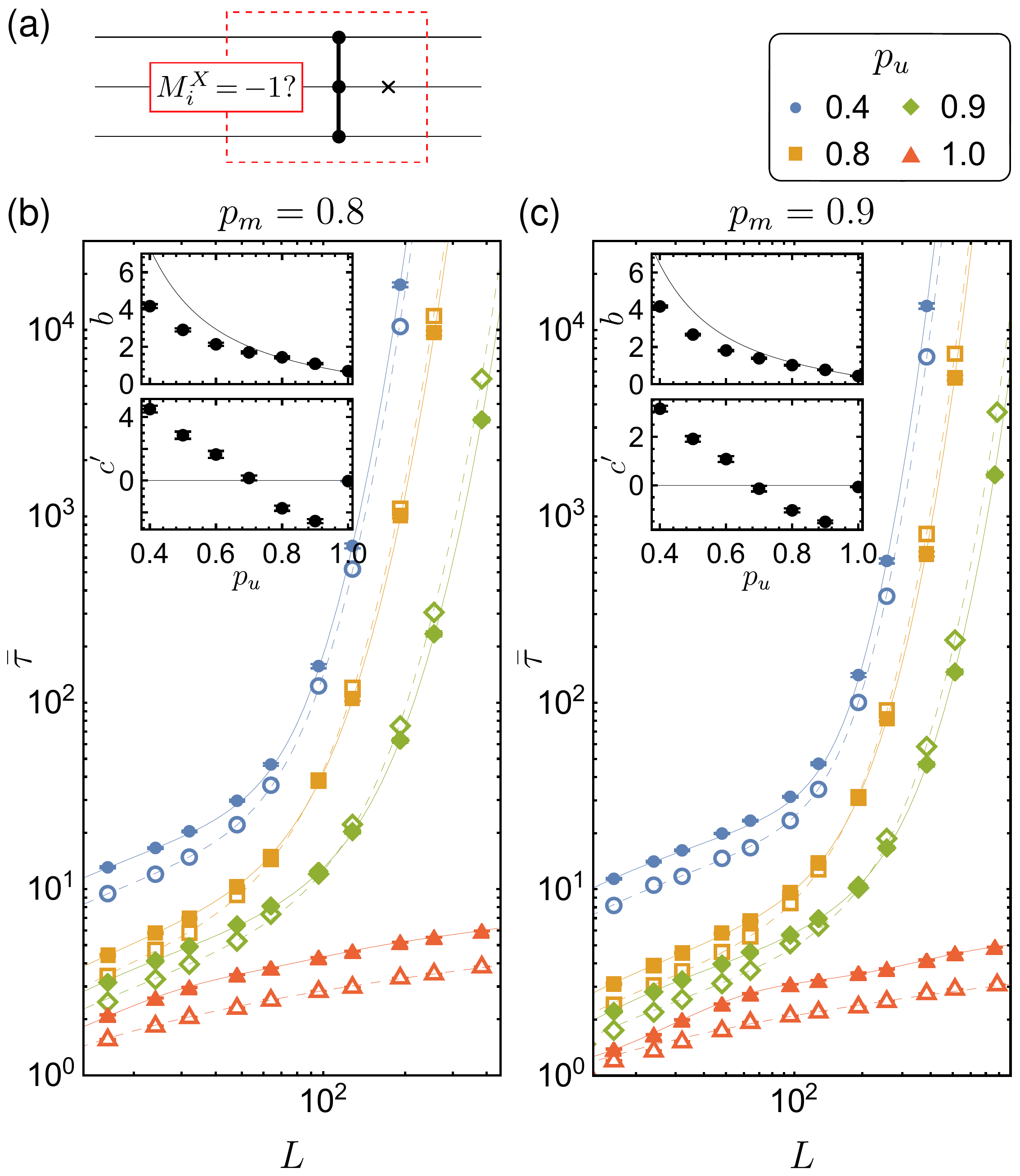}
  
  \caption{(a)~Local decoder that is conditioned upon the result of the measurement outcome: if the result is $M^X_i = - 1$, then the decoder is used. Unitaries and measurements in the decoder are applied with probabilities $p_u$ and $p_m$ respectively. (b,c)~Mean time $\bar\tau$ of achieving a cat state in a stochastic circuit with a decoder (open markers) and without the decoder (filled markers) for (b)~$p_m = 0.8$, and (c)~$p_m=0.9$. Dashed lines show fits to Eq.~(\ref{eq:mean-time-generic}). The insets show the best-fit parameters $b$ and $c'$, where $c'=(c-p_X)\times 10^3$. In insets for $b$, solid lines show corresponding coefficients without the decoder [Eq.~(\ref{eq:mean-time-zz})].}
  \label{fig:decoder_results}
\end{figure}

In our case, we introduce a local decoder in the state preparation protocol circuit in an attempt to impact the dynamics locally, as shown in Fig.~\ref{fig:decoder_results}(a).
The decoder is conditioned upon the result of the measurement from the previous layer (if no measurement has been applied, then the outcome is assumed to be random): if the result is $M^X_i=-1$, the decoder is applied, otherwise it is not applied. The decoder is composed of two neighboring unitaries followed by a measurement in the $X$ direction. Each part of the decoder is assumed to be governed by the same $p_u$ and $p_m$ as in the full circuit (e.g.\@ the decoder is imperfect in our simulations).

\changes{The form of the decoder is chosen due to the initial all-minus state. The rationale is to detect and rectify errors early in the circuit, while the signatures of the initial state are still identifiable. To achieve this, we investigate how the $M^X_i=-1$ result correlates with the presence of errors when a 3-qubit circuit is applied to the all-minus state. A careful consideration (see Table~\ref{tab:local-circuits} in Appendix~\ref{sec:appendix-exp-time}) shows that the decoder will change the probability of obtaining a $Z_i Z_{i+2}$ stabilizer when starting from an all-minus state.} More specifically, if no decoder is applied, this probability is $p_u^2 p_m$, but (for simplicity) if a perfect decoder is applied, the probability increases to $p_u^2 p_m + (1 - p_u)^2 p_m + 0.5 (1 - p_u)^2 (1 - p_m)$ (corresponding to the first, fourth and the eighth rows of Table~\ref{tab:local-circuits}). An imperfect decoder considered in this section will also increase the probability, albeit by a smaller value.

Numerical results [see Fig.~\ref{fig:decoder_results}(b,c), open markers] indicate that indeed the decoder takes care of recovering the $Z_i Z_{i + 2}$ symmetries faster, however, the exponential behavior for large system sizes persists. To better understand the change, we fit the results to the form
\begin{equation}
  \bar\tau = a + b \log(L/2) + c^{-L/2},
  \label{eq:mean-time-generic}
\end{equation}
with the corresponding best-fit parameters shown in the insets in Fig.~\ref{fig:decoder_results}(b,c). The fitting ansatz includes a logarithmic term similar to that of Eq.~(\ref{eq:log-mean-time}) and an exponential term, where the exponent base $c < 1$ serves the same purpose as probability $p_X$ in Eq.~(\ref{eq:mean-time-Z2}). The coefficient of the log term,  $b$ is reduced by the decoder, as compared to the protocol without the decoder [solid lines, Eq.~(\ref{eq:mean-time-zz})]. However, we find that the base $c$ of the exponential term can sometimes be increased (for low $p_u$), indicating a slower exponential growth, but for large values of $p_u$ we find $c$ is lowered, and the exponential growth is even faster than in the protocol without the disorder. Therefore, we conclude that the proposed decoder works well for low values of $p_u$, or for small systems when the logarithmic term dominates. The exponential times, however, cannot be impacted substantially by local decoding.

\subsection{Halting protocol}
\label{sec:halting-protocol}

\begin{figure}[tb]
  \centering
  \includegraphics[width = 0.99\columnwidth]{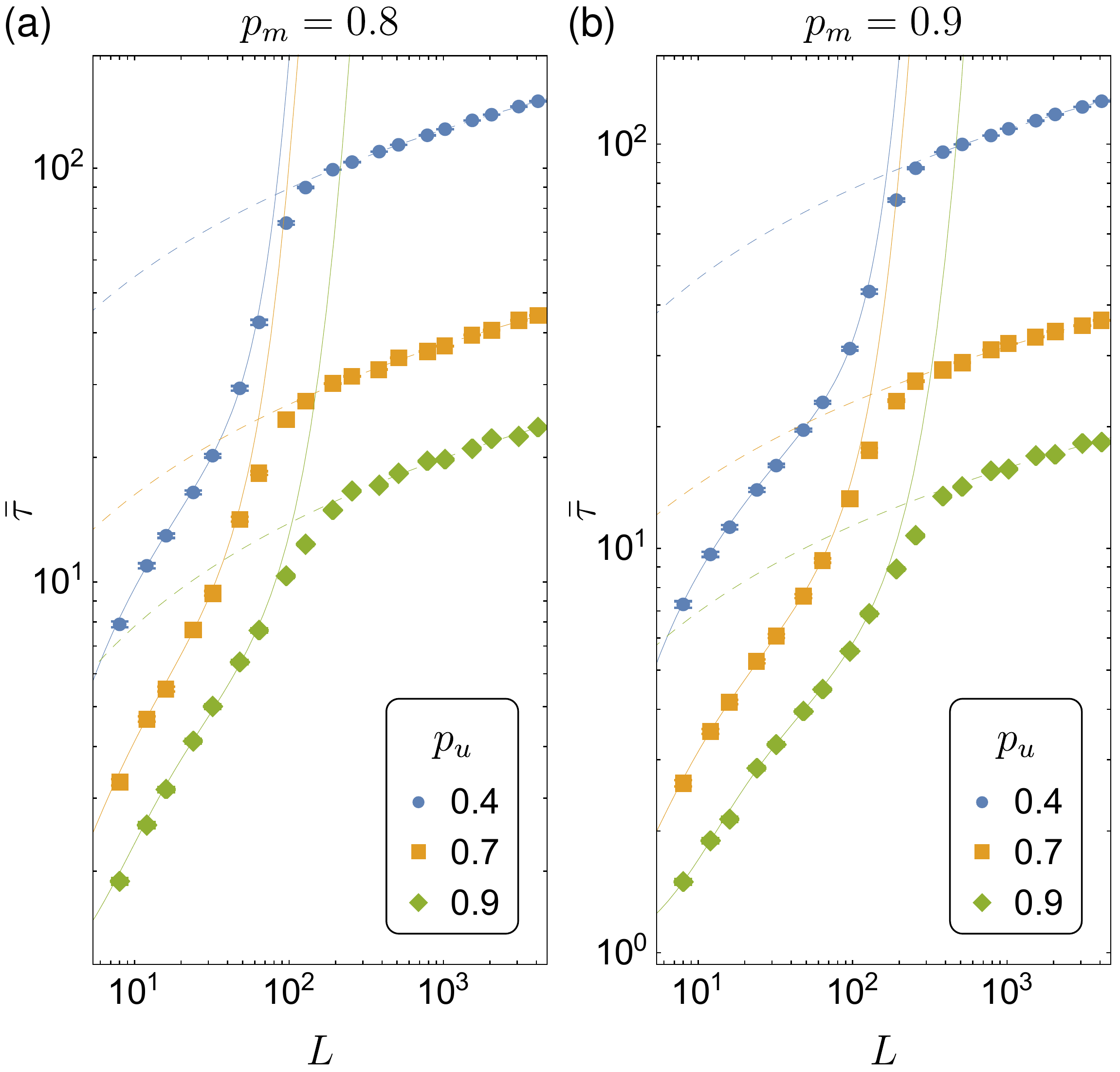}
  
  \caption{Mean time $\bar\tau$ of achieving a cat state in a protocol where the unitary evolution is halted. Solid lines are analytical predictions based on Eqs.~(\ref{eq:CDF_ZZ}) and (\ref{eq:mean-time-Z2}) for the case without halting. Dashed lines are fits to the numerical data for large system sizes, with the fitting form $\bar\tau = a+b\log(L)$.}
  \label{fig:halting_results}
\end{figure}

Having the knowledge of the two timescales, we can construct another speedup protocol for our stochastic state preparation scheme, where we aim to impact the global properties of the circuit. First, we estimate the moment when the $Z_i Z_{i+2}$ stabilizers are reached from our analytical results for $\bar\tau_\ZZ$, specifically Eq.~(\ref{eq:time-fidelity}) involving fidelity. We then turn off the unitary evolution, and the subsequent applications of measurements will quickly lead to the recovery of the $\mathbb{Z}_2$ symmetry. Note that by stopping the unitary evolution we essentially impact the system globally by altering the original protocol.

In detail, we halt the unitary evolution after times $[2 \tau(\phi,p_u^2 p_m) + 4]$, where the factor of $2$ accounts for the even-odd-layer effects close to $p_u = 1$, and the additive constant helps to describe the limit of $\{p_u = 1, p_m = 1\}$ correctly. We set the fidelity to $\phi=99\%$, which makes sure that in at least $99\%$ of random realizations, the circuit has reached the $Z_i Z_{i+2}$ stabilizers. Numerical results in Fig.~\ref{fig:halting_results} confirm that the halting of the unitary process reduces the time $\bar\tau$ from exponential in the system size (solid lines) to logarithmic (dashed lines). We also confirm that the probability of failure is below $1\%$, as suggested by the analytics.

\subsection{Forced global symmetries}

\begin{figure}[tb]
  \centering
  \includegraphics[width = 0.99\columnwidth]{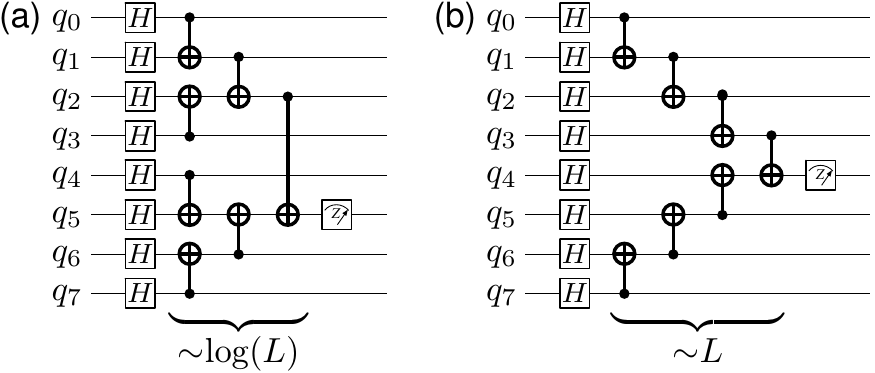}
  
  \caption{Possible ways to realize $\prod_i X_i$ measurements using Hadamard gates, CNOT gates and single qubit Z\mbox{-}measurements. (a)~Allowing nonlocal two-site gates leads to $\sim \log(L)$ circuit depth, while (b)~local two-site gates lead to extensive $\sim L$ depth.\label{fig:circuit-prodx}}
\end{figure}

Another speedup protocol would be to enforce the global symmetry of $\prod_{i\text{ odd}} X_i$ at the end of the evolution. One way to do it is to measure the corresponding nonlocal operator, which is equivalent to performing a global parity check, a process often used in quantum error correction~\cite{Gottesman1998, Terhal2015}. This, however, is experimentally hard to implement~\cite{Saira2014, Kelly2015, Blumoff2016}, as one would need to use at least a $\log(L)$-depth circuit with \textit{long-range} two-site unitaries and one-site measurements [see Fig.~\ref{fig:circuit-prodx}(a)], or an $L$-depth circuit if only \textit{local} two-site unitaries and one-site measurements are allowed [see Fig.~\ref{fig:circuit-prodx}(b)] (the latter, due to a requirement of locality, is directly related to the light cone). Assuming the application of each gate is imperfect, this would lead us back to exponential times and is therefore unfeasible. On the other hand, recent advances in quantum computing architectures show promise of fault-tolerant multi-qubit parity-check measurements~\cite{Negnevitsky2018, Hilder2022, Ni2023}, which would make the enforcement of global stabilizer a more viable option.

\section{Stability of the long-range entanglement in the stochastic protocol}
\label{sec:stability}

In this section, we study how different perturbations to the time evolution operator affect the existence of the LRE state. Specifically, we consider timing imperfections and additional terms in the unitary evolution, such as a transverse field term and a transverse interaction term.

\subsection{Timing imperfections}

We investigate the effects of imperfect timing in the unitary evolution by setting $\Delta t = (\pi / 4)\, \theta$ in Eq.~(\ref{eq:unitary}). Here, $\theta$ is not necessarily equal to 1, which means the state is no longer always a stabilizer state, requiring dense representation \changes{(a state being defined through $2^L$ complex coefficients of the computational basis~\cite{Sandvik2010})} and allowing only for small system sizes. To determine whether the desired state is achieved, we calculate expectation values of desired stabilizers $|\langle Z_i Z_{i+2} \rangle|$, $|\langle \prod_{i\text{ odd}} X_i \rangle|$, and the mutual information between two antipodal unmeasured sites A and B,
\begin{equation}
    I_2(A:B) = S_\text{A} + S_\text{B} - S_\text{AB}.
\end{equation}
When the state becomes a cat state, the two expectation values reach 1, while $I_2$ reaches $\log(2)$. Finally, we average the results over many trajectories (denoted by overbar).

\begin{figure}[tb]
  \centering
  \includegraphics[width = 0.99\columnwidth]{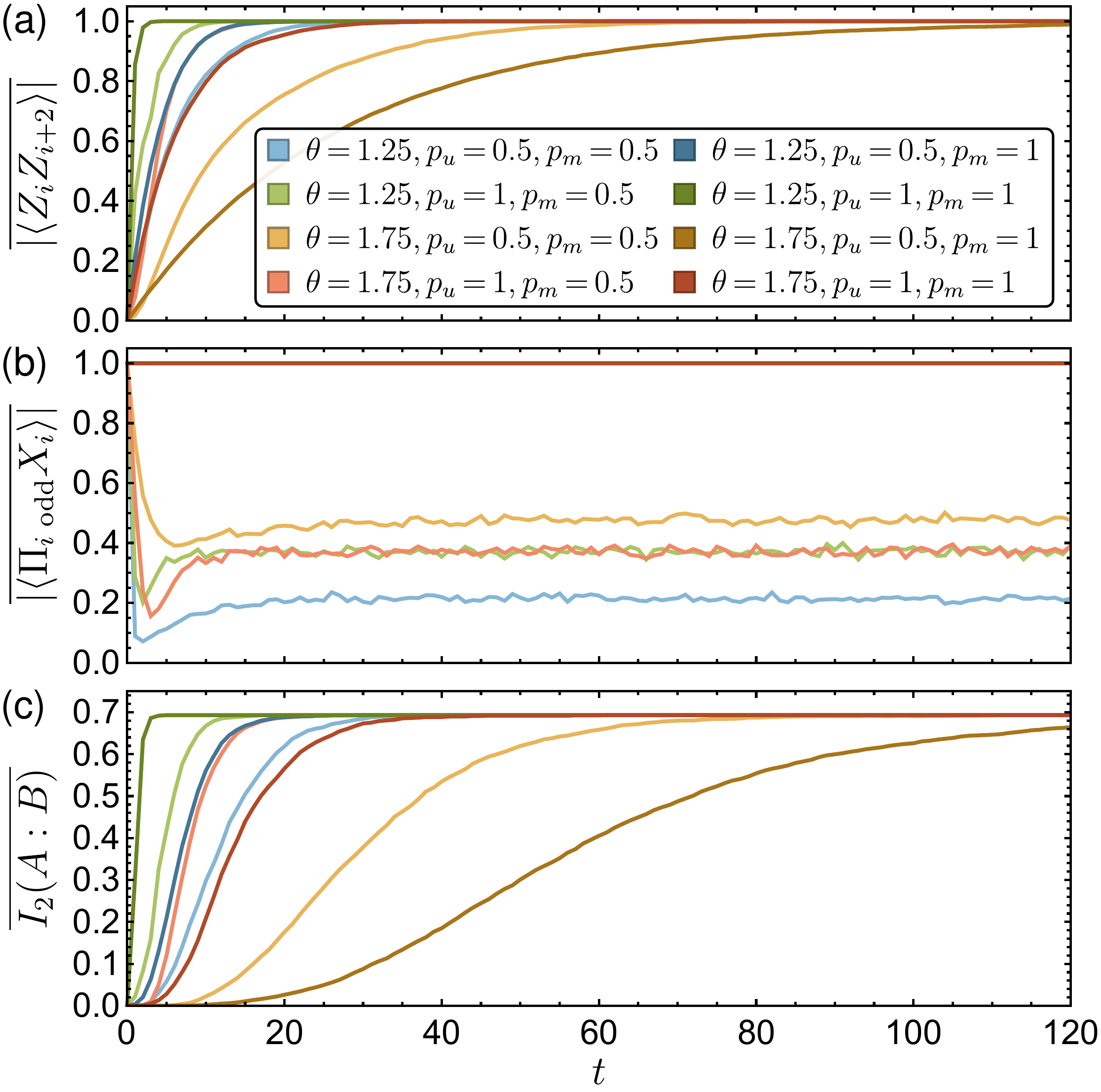}
  \caption{Protocol with time imperfections: averaged expectation value of (a)~stabilizers $|\langle Z_i Z_{i+2} \rangle|$, (b)~the global symmetry $|\langle \prod_{i\text{ odd}} X_i \rangle|$, and (c)~the averaged mutual information between two antipodal unmeasured sites A and B, $I_2(A:B)$. Legend in (a) applies to (b) and (c). The system size is $L=16$.
  \label{fig:time-imperfections}
  }
\end{figure}

Numerical results for $L=16$ shown in Fig.~\ref{fig:time-imperfections} reveal that the cat state is recovered at long times when $p_m = 1$, and $p_u$ and $\theta$ are arbitrary. The times get longer when $\theta$ is increased from 1 to 2 and when $p_u$ is decreased, which is to be expected. However, when $p_m < 1$, we observe that the $Z_i Z_{i+2}$ stabilizers are recovered, but the $\mathbb{Z}_2$ symmetry of $\prod_{i \text{ odd}} X_i$ is not present. This is a qualitatively similar result to those when no time imperfections are present. We expect that this behavior is governed by similar processes in both cases, i.e.\@ for $p_m=1$, the measurements on all sites are responsible for fixing the global $\prod_{i \text{ odd}} X_i$ symmetry.

Thus, we conclude that the time imperfections have surprisingly little impact on the qualitative behavior of the circuit. In fact, one can show that $|\langle Z_i Z_{i+2} \rangle|$ never decreases on average after applying a one-time-step circuit on a generic state (see Appendix~\ref{sec:appendix-time-imperfections-proof}). Together with the fact that $Z_i Z_{i+2}$ stabilizers are stable under the evolution with timing imperfections, this implies that the local stabilizers of the cat state are a fixed point of this process.

\subsection{The appearance of a measurement-induced entanglement transition}

Let us now consider unitary perturbations in the form of additional terms in the Hamiltonian in Eq.~(\ref{eq:unitary}). First, we investigate a situation where a transverse field term is introduced, and the Hamiltonian becomes $H = \sum_i Z_i Z_{i + 1} + \Gamma_X \sum_i X_i$. For generic values of the field strength $\Gamma_X$, the stabilizer formalism can no longer be used, and instead, we employ dense methods. We find (see Appendix~\ref{sec:appendix-stability-transverse-field} for more details) that the transverse field generically leads to a steady state that is not a cat state, therefore destroying long-range entanglement.

\begin{figure*}[tb]
  \centering
  \includegraphics[width = 0.99\textwidth]{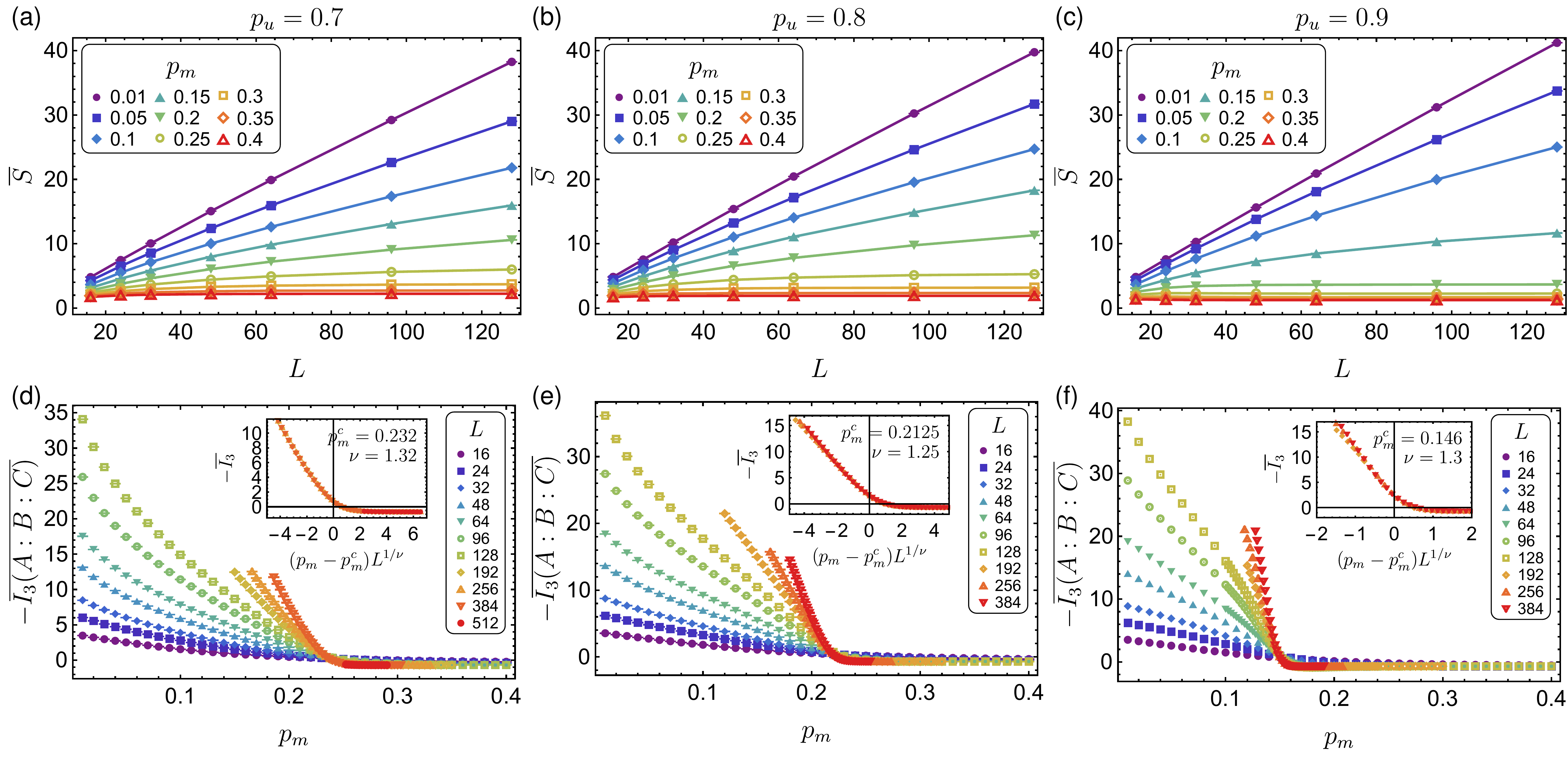}
  
  \caption{Measurement-induced transition when the unitary evolution includes both ZZ and XX terms. (a,b,c) Average half-chain entanglement entropy $S$ as a function of the system size. A clear pattern emerges, where for low $p_m$ we observe the volume law $S \sim L$ and for high $p_m$ we see the area law $S \sim O (1)$. (d,e,f) Tripartite mutual information $I_3(A:B:C)$ can be used as a transition diagnostic. Insets show data collapses of $I_3$ and the best-fit values of critical parameters.
  The left column is for $p_u=0.7$, the middle column is for $p_u=0.8$, and the right column is for $p_u=0.9$.
  \label{fig:XX-ZZ-miet}
  }
\end{figure*}

We proceed with considering an additional interaction term, which preserves the Clifford character of the circuit, as a perturbation in the unitary element of the protocol. When the unitary evolution is given by Eq.~(\ref{eq:unitary}) without any extra terms, the entanglement entropy grows up to a system-size independent constant value. Therefore, the steady state of such evolution may be long-range entangled, yet never reach a volume-law entangled state. In order to investigate the robustness of this feature, we investigate a model where the evolution operator is replaced by
\begin{align}
  U & = \prod_i \exp \left[ - i \frac{\pi}{4} (X_i X_{i + 1} + Z_i Z_{i + 1})
  \right] \\
  & = \prod_i \left[ \exp \left( - i \frac{\pi}{4} X_i X_{i + 1} \right) \exp \left( - i \frac{\pi}{4} Z_i Z_{i + 1} \right) \right] \\
  & = \prod_i U_{i,i+1},
  \label{eq:unitary-miet}
  \nonumber
\end{align}
where the two-site unitaries $U_{i,i+1}$ are applied in a brick-work fashion and do not commute with each other [as opposed to the unitaries in Eq.~(\ref{eq:unitary})]. This translates to adding the $\sum_i X_i X_{i+1}$ interaction term in the Hamiltonian $H$ from Eq.~(\ref{eq:unitary}). The unitary $U_{i,i+1}$ is a Clifford gate and preserves stabilizer states, allowing us to use the tableau formalism. We are interested in the steady-state properties, hence we first evolve the system for equilibration time proportional to the system size, before considering the steady-state values.

We find that this evolution operator $U$ induces growth of entanglement, and consequently the emergence of a \textit{measurement-induced entanglement transition}~\cite{Li2018, Skinner2019, Chan2019, Szyniszewski2019} within the steady-state dynamics of the system. This novel type of transition occurs when the entropy changes its behavior from extensive (volume law) to sub-extensive (area law) when measurement frequency $p_m$ is varied. The volume law phase occurring for infrequent measurements is expected to be useful for quantum error correction~\cite{Li2021, Choi2020}, although the existence of an efficient decoder appears non-trivial.

When $p_u < 1$, we find clear signatures of a measurement-induced transition within the entanglement entropy [see Fig.~\ref{fig:XX-ZZ-miet}(a-c)]: for low values of $p_m$, the entropy is extensive, $S \sim L$, while for high values of $p_m$, the entropy saturates to a constant value for large system sizes, $S \sim O (1)$. In order to pinpoint the transition and its critical properties, we use tripartite mutual information~\cite{Zabalo2020} given by
\begin{align}
  I_3 (A : B : C) & = S_{\text{A}} + S_{\text{B}} + S_{\text{C}} \nonumber \\
  & \quad - S_{\text{AB}} - S_{\text{BC}} - S_{\text{AC}} + S_{\text{ABC}},
\end{align}
where the system is divided into four equal continuous regions A, B, C, D, each of length $L / 4$. This quantity serves as a proper transition diagnostic, as it cancels the boundary contributions to the entropy~\cite{Kitaev2006, Zabalo2020}. Near the transition region, $I_3$ obeys the following scaling relation~\cite{Zabalo2020},
\begin{equation}
  I_3 \sim F [(p_m - p_m^c) L^{1 / \nu}],
  \label{eq:I3-scaling}
\end{equation}
where $F [\cdot]$ is an unknown scaling function, $p_m^c$ is the critical point, and $\nu$ is the critical exponent of the correlation length. Fig.~\ref{fig:XX-ZZ-miet}(d-f) shows the results for $I_3$, while the insets show the data collapses for the scaling ansatz of Eq.~(\ref{eq:I3-scaling}), together with best-fit values for $p_m^c$ and $\nu$. Albeit the critical point $p_m^c$ is not universal, we expect that $\nu$ should be universal and independent of the value of $p_u$ chosen. Indeed, our estimates of $\nu \approx 1.3$ are roughly constant for $p_u \in \{ 0.7, 0.8, 0.9 \}$. Additionally, this value is consistent with the corresponding exponent found in the literature for random stabilizer circuits with measurements, $\nu = 1.265 (15)$~\cite{Sierant2022, Gullans2020, Zabalo2020, Lunt2021}, which is known to be in the universality class of perturbed Potts model~\cite{Li2021}.

\begin{figure}[tb]
  \centering
  \includegraphics[width = 0.99\columnwidth]{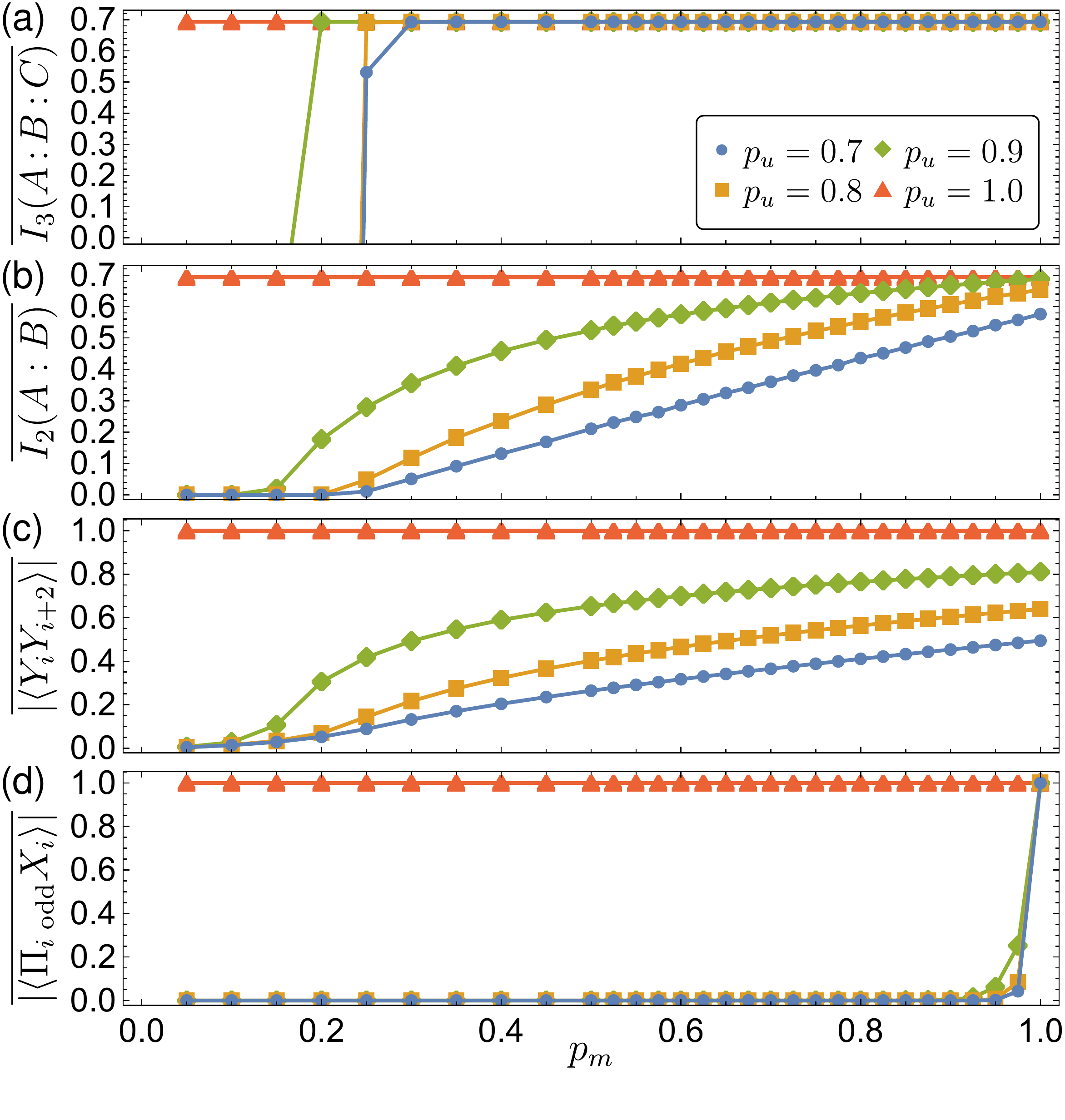}
  
  \caption{Proximity of the steady state to a cat state in the $Y$ direction for a circuit with unitary evolution including both ZZ and XX interaction terms, system size $L = 384$. (a)~Tripartite mutual information $I_3$ is nearly $\log(2)$ in the area law, while being mostly negative in the volume-law phase. (b) Bipartite mutual information $I_2$ vanishes in the volume law and is finite in the area law. (c) Expectation value of local stabilizers of the cat state, $|\langle Y_i Y_{i+2} \rangle|$, vanishes in the volume law and is finite in the area law. (d)~Expectation value of the global stabilizer of the cat state, $|\langle \prod_{i\text{ odd}} X_i\rangle|$, vanishes nearly everywhere in the phase diagram.
  \label{fig:XX-ZZ-miet-cat}
  }
\end{figure}

We also note that, contrary to the random models, we find that $I_3$ becomes positive in the area law phase, which implies that information about correlations in the system is shared between different parts of the system~\cite{Seshadri2018, Iyoda2018, Rangamani2015, Rota2016} (i.e.\@ mutual information $I_2$ is not monogamous). This is somewhat expected, as random states generically have non-positive $I_3$~\cite{Rangamani2015}, while certain long-range entangled states (such as the GHZ state) have positive $I_3$~\cite{Rota2016}. In fact, for $p_u=1$, we find that the circuit always produces a cat state in the $Y$ basis, with $I_3 = \log(2)$. Numerical results for $L=384$ in Fig.~\ref{fig:XX-ZZ-miet-cat} show that in the area-law phase, $I_3$ is remarkably stable at $\log(2)$, implying similar properties to a cat state. We also find that in the area law and near $p_u\sim 1$, the state has a high chance of recovering the $Y_i Y_{i+2}$ local symmetry, albeit, below $p_u = 1$ the $\mathbb{Z}_2$ global symmetry is generically not present.

The presence of the $Y_i Y_{i+2}$ stabilizers implies a possible existence of a cat state in the $Y$ direction. In fact, as we increase the system size, we find that the kink in Fig.~\ref{fig:XX-ZZ-miet-cat}(c) near the transition becomes more pronounced, leading us to conclude that in the volume-law phase, the chance of obtaining a cat state vanishes quickly as compared to the area-law phase. This can be intuitively understood by noticing that the volume law is ergodic -- the state explores the whole Hilbert space roughly equally, with a finite but rapidly diverging time in system size to achieve a cat state ($\bar\tau \sim 2^{L^2/2}$, which is the number of possible stabilizer states of $L$ qubits~\cite{Aaronson2004}). On the other hand, in the area law, the $Y_i Y_{i+2}$ stabilizers appear with finite probability, and one only requires to wait until the $\mathbb{Z}_2$ symmetry is recovered.

To summarize, when including interaction terms with nearest neighbor isotropic $XX$ and $ZZ$ terms, the system exhibits a measurement-induced entanglement transition. The state is featureless in the volume law, while in the area-law phase, it behaves similarly to a cat state in the $Y$ basis when the probability of applying the unitaries approaches unity.

\section{Long-range entanglement in higher dimensions}
\label{sec:2d}

Higher-dimensional systems are of vast importance for state preparation protocols, as they host a plethora of exotic phases, such as spin liquids and topological order~\cite{Kitaev2006toric}. Therefore, the move from one to higher dimensions is a non-trivial step, as it offers new avenues for quantum error-correcting codes. Specifically, although LRE states in 1D, such as cat states, are important, they offer limited use in quantum algorithms. On the other hand, the 2D toric code state~\cite{Kitaev2003, Kitaev2006toric} serves as a paradigmatic example of topological quantum error correction which has recently been realized in Rydberg atom arrays~\cite{Semeghini2021, Bluvstein2022} and a superconducting quantum processor~\cite{Satzinger2021}. This gives hope that realizations of topological quantum computing through surface codes~\cite{Dennis2002, Fowler2012, Erhard2021} could be possible in the near future. Recent advancements~\cite{Andersen2022, Iqbal2023feedforward} also lead us to consider anyonic excitations present in the toric code as viable proposals for fault-resistant topologically-protected quantum computing.

\begin{figure}[t]
  \centering
  \includegraphics[width = 0.99\columnwidth]{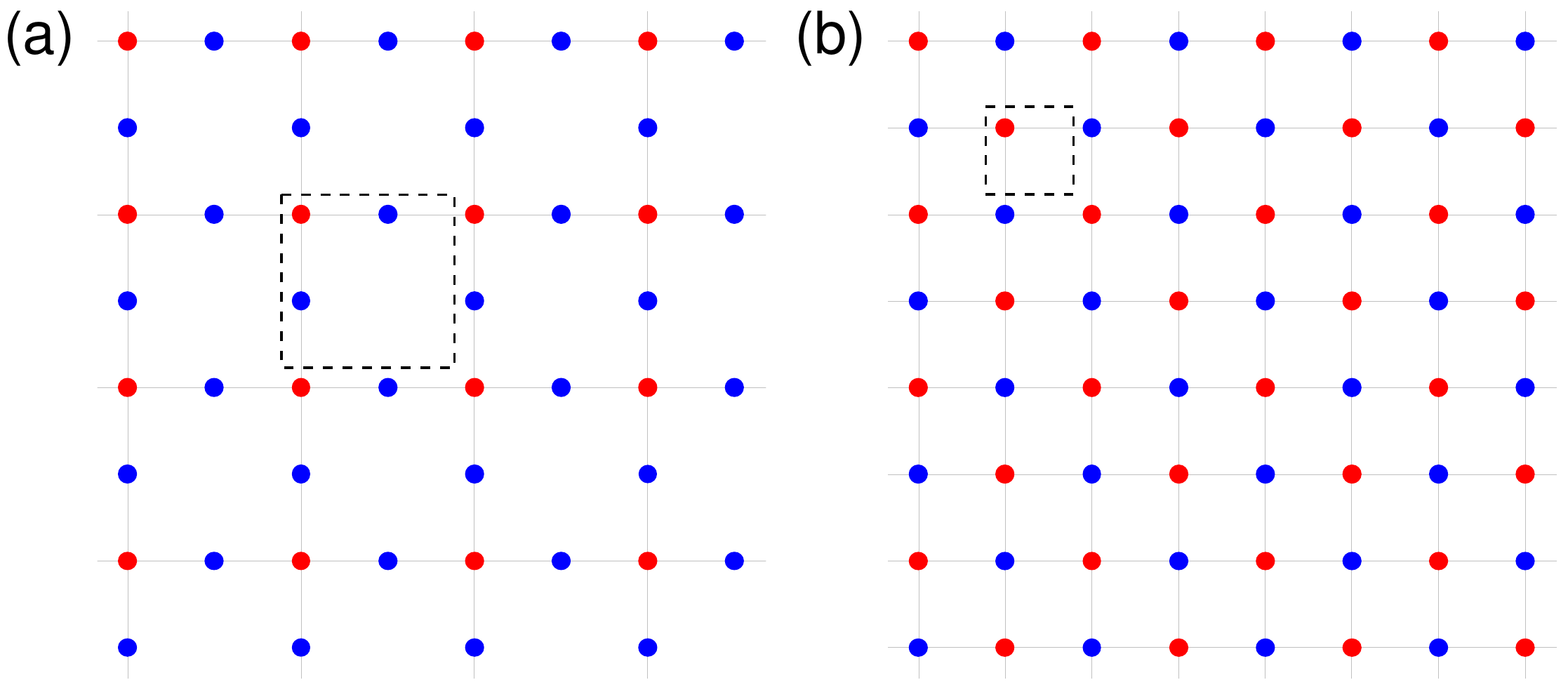}
  \caption{(a) Lieb lattice. (b) Square lattice. We perform unitary operations between sites connected by a line, and measurements on sites marked by red dots. Dashed regions designate unit cells.
  \label{fig:2d-lattices}}
\end{figure}

In this section, we consider two-dimensional lattices subject to the stochastic protocol that we introduced in 1D, details of which we describe below. Non-stochastic state preparation protocols with measurements were already shown to be effective in producing a toric code state~\cite{Raussendorf2005}. Specifically, we will discuss the results for the Lieb lattice and the square lattice, shown in Fig.~\ref{fig:2d-lattices}, with periodic boundary conditions putting the system on a torus. Unitary gates from Eq.~(\ref{eq:unitary}) are now applied between each pair of neighboring sites (solid lines in the figure), and the measurements in the $X$ direction are applied on red sites only. $L$ now designates the linear size of the lattice, i.e.\@ the number of unit cells (dashed regions) in either $X$ or $Y$ direction. Therefore, for the Lieb lattice, the total number of spins is $3 L^2$, while the square lattice consists of $L^2$ spins. On the Lieb lattice, it is useful to define a ``star'', which is composed of one red site surrounded by four blue sites, and a ``plaquette'', which is composed of four blue sites with no site in the middle. This introduces a standard language for the description of the toric code states.

\subsection{Toric code states on the Lieb lattice}

\begin{figure}[t]
  \includegraphics[width = 0.99\columnwidth]{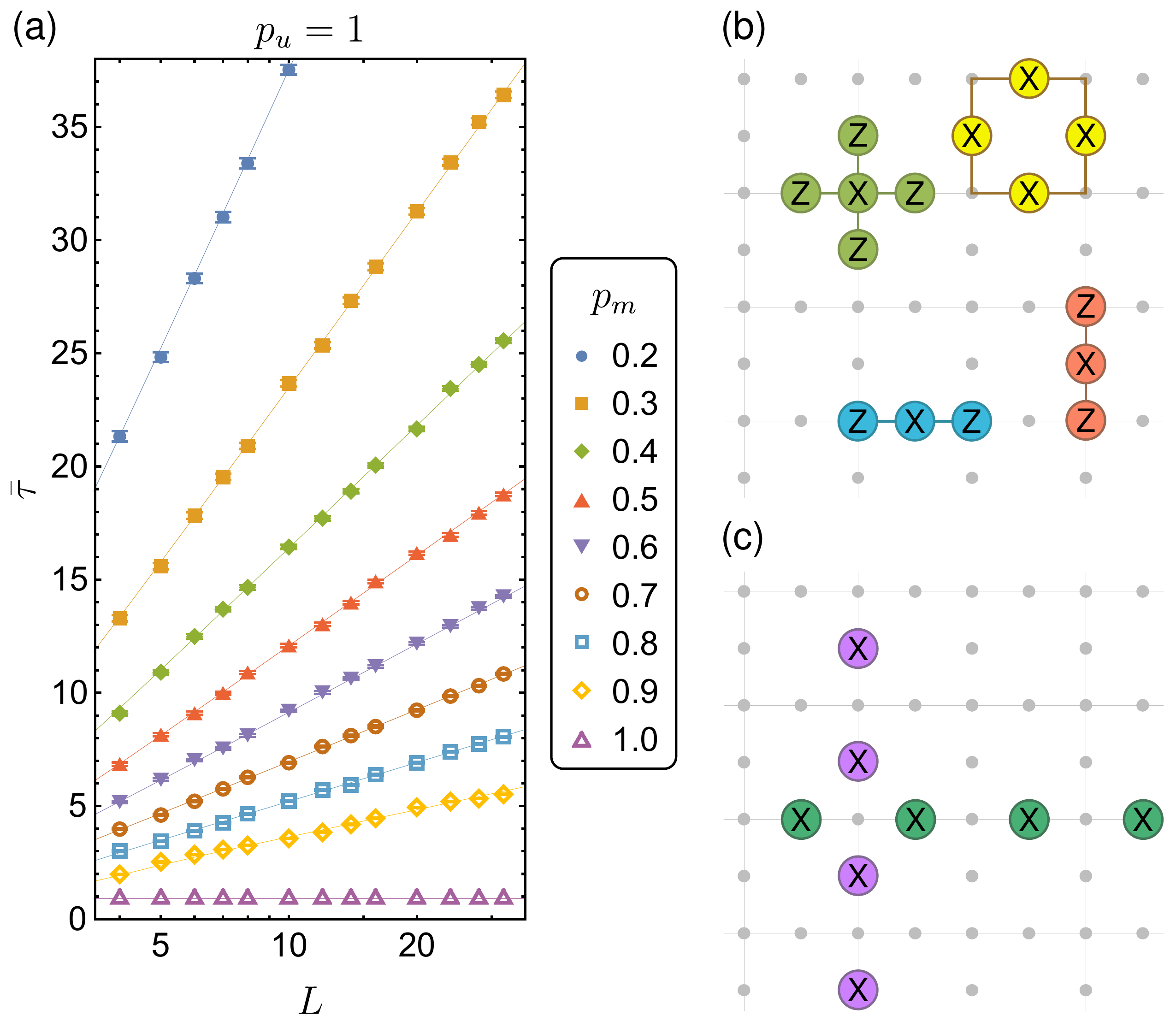}
  \caption{Stochastic protocol results on a Lieb lattice. (a)~Mean time to achieve a toric code state on a Lieb lattice. (b)~Local stabilizers of the cluster state: the star stabilizer (in green), horizontal plaquette edge stabilizer (in blue), and vertical plaquette edge stabilizer (in red). Plaquette edge stabilizers can be combined to obtain a plaquette stabilizer (in yellow). (c)~Global stabilizers of the cluster state are vertical and horizontal line stabilizers (in purple and in dark green, respectively).
  \label{fig:2d-lieb-results}}
\end{figure}

We start the discussion of the state preparation protocol on a Lieb lattice by considering the exact case of $p_u=1$ and $p_m=1$, where after one layer of unitaries, the system is in a cluster state. Local stabilizers of the cluster state on the Lieb lattice are shown in Fig.~\ref{fig:2d-lieb-results}(b), namely a star stabilizer $X_{i,j} Z_{i-1,j} Z_{i+1,j} Z_{i,j-1} Z_{i,j+1}$ (in green), and two versions of the stabilizer ${ZXZ}$ placed on the edge of the plaquette horizontally or vertically (in blue and red). Four ${ZXZ}$ stabilizers around a plaquette can be combined to obtain a plaquette stabilizer $B_p := X_{i+1,j} X_{i,j+1} X_{i+1,j+2} X_{i+2,j+1}$ (in yellow). Global stabilizers can be obtained by combining the ${ZXZ}$ stabilizers in either horizontal or vertical lines, giving either $\prod_{i} X_{2i+1,j}$ or $\prod_{j} X_{i,2j+1}$, which are usually called line symmetries [see Fig.~\ref{fig:2d-lieb-results}(c)].

Measurements have the following effect: the ${ZXZ}$ stabilizers disappear, while the star stabilizers, plaquette stabilizers, and global symmetries are left intact. Since the measurements produce stabilizers $\pm X_{i,j}$ on the measured sites, these can be combined with the star stabilizers, obtaining the stabilizers $\pm A_v := \pm Z_{i-1,j} Z_{i+1,j} Z_{i,j-1} Z_{i,j+1}$, which we will call the toric code vertex stabilizers. The resulting state on the blue sublattice is an eigenstate of the toric code Hamiltonian~\cite{Kitaev2003, Kitaev2006toric},
\begin{equation}
  H = - \sum A_v - \sum B_p,
  \label{eq:toric-code-hamiltonian}
\end{equation}
a well-known model exhibiting topological protection of qubits.

Similar to the previous protocol for the cat state, if one lowers $p_u$ and/or $p_m$, the toric code state is still achieved, but within some mean time $\bar\tau$. The results for $p_u = 1$ and varying $p_m$ are shown in Fig.~\ref{fig:2d-lieb-results}(a), where we find that the mean time is logarithmic with the system size. This can be explained as follows. The $B_p$ stabilizers are always present after the cluster state is created at $t=1$. Each measurement may create (with probability $p_m$) a $\pm A_v$ stabilizer if it is in the odd layer (cf.~Sec.~\ref{sec:cat-pu1}), which is then stable towards further evolution. Finally, $(L^2-1)$ $A_v$ stabilizers are needed to create all toric code vertex stabilizers, as the last one follows naturally by combining all others. Therefore, the mean time is
\begin{equation}
  \bar\tau \approx 2 \left( \frac{\log (1/L^2)}{\log (1 - p_m)} - \frac{\gamma - 1}{\log (1 - p_m)} \right),
\end{equation}
which coincides with our numerical data [see solid lines vs markers in Fig.~\ref{fig:2d-lieb-results}(a)].

\begin{figure}[tb]
  \includegraphics[width = 0.99\columnwidth]{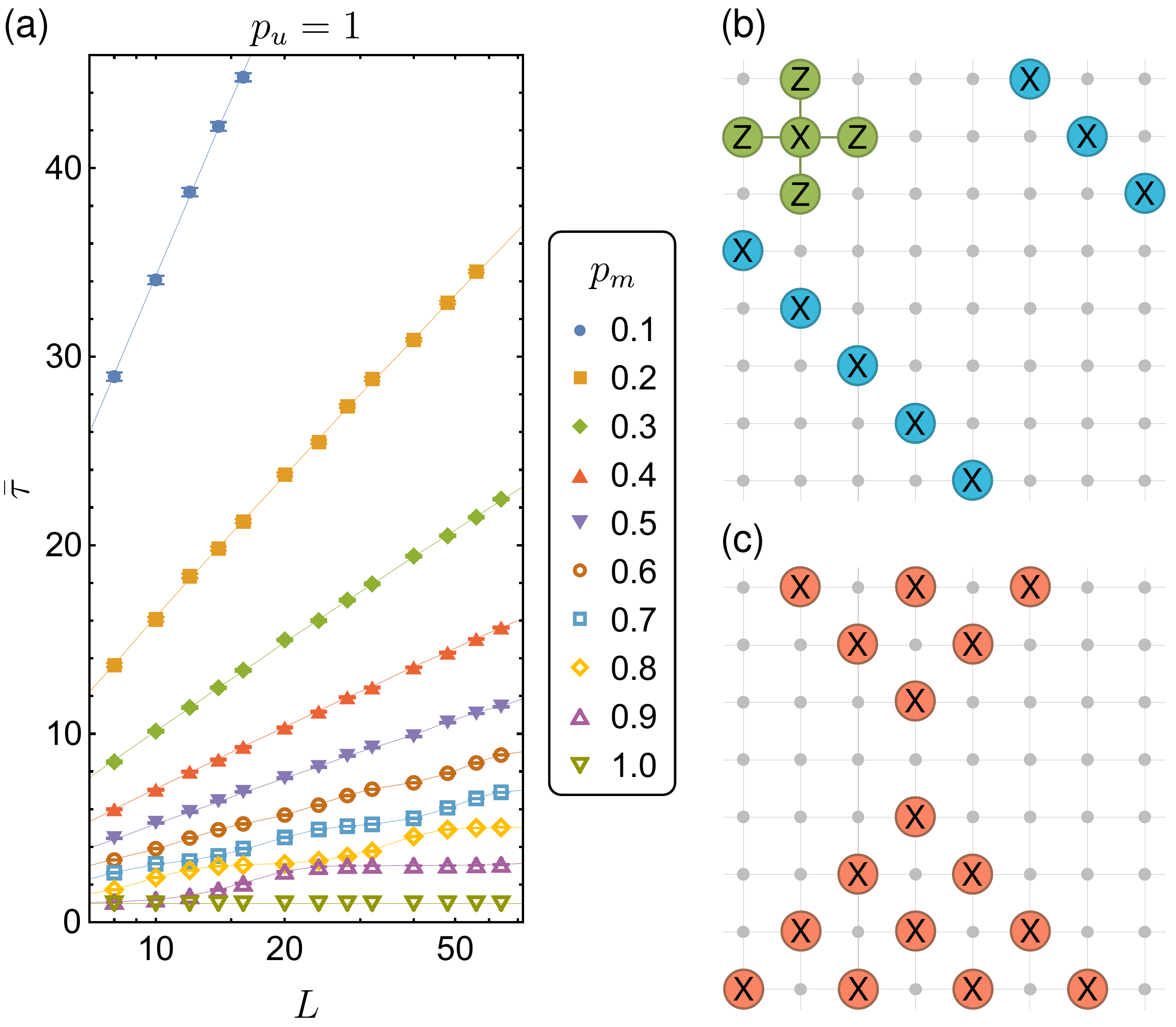}
  \caption{Stochastic protocol results on a square lattice. (a)~Mean time to achieve a Xu-Moore state. (b)~The star stabilizer (in green) is a local stabilizer of the cluster state on the square lattice, and can be combined to get a global diagonal stabilizer (in blue). (c)~Other global stabilizers have a cone shape (in red).
  \label{fig:2d-square-results}}
\end{figure}

When $p_u<1$, we find that the time to reach a toric code state grows rapidly, owing to the fact that the plaquette stabilizers $B_p$ are no longer stable. We expect this growth to be exponential, as each $B_p$ has a probability of occurrence approximately equal to $p_m p_u^4$, i.e.\@ the mean time is $\sim 1/(p_m p_u^4)^{L^2}$. This time can be cut down by applying a halting protocol, as in Sec.~\ref{sec:halting-protocol}, where only a logarithmic time will be needed to reach a toric code state.

The toric code states achieved in this stochastic protocol are not necessarily ground states -- instead, generically, the protocol produces excited toric code states, which can be viewed as quasiparticles obeying anyonic statistics. The resulting state consists of defects connected by well-defined strings of flipped spins, as opposed to a superposition of all possible paths between the two endpoints. During the braiding of two anyons, this gives direct access to the process where the quasiparticle acquires a phase factor. Therefore, our protocol opens an avenue for further investigations of multi-particle anyon braiding. Deforming the structure of the Lieb lattice can also lead to non-abelian anyons~\cite{Bombin2010}, which are of particular interest for topological quantum computation~\cite{Freedman2001, Andersen2022, Iqbal2023}.

\subsection{Xu-Moore states on the square lattice}

We turn to the discussion of the protocol on a square lattice. First, we again consider the exact case of $p_u=1$ and $p_m=1$. After the unitary evolution, the resulting state is a cluster state on a two-dimensional square lattice, which can be described by local star stabilizers $X_{i,j} Z_{i-1,j} Z_{i+1,j} Z_{i,j-1} Z_{i,j+1}$ [in green in Fig.~\ref{fig:2d-square-results}(b)]. These can be combined alongside any diagonal to get global diagonal line stabilizers of $\prod_i X_{i,i+c}$ or $\prod_i X_{i,-i+c}$ [in blue in Fig.~\ref{fig:2d-square-results}(b)]. The local star stabilizers can also be combined to produce global cone stabilizers [in Fig.~\ref{fig:2d-square-results}(c)], which have a cone shape on an infinite lattice, while on a finite periodic lattice resemble a diamond.

All aforementioned global symmetries survive the application of measurements. The star stabilizers centered around the red sites also survive, and can be combined with the new measurement stabilizers $\pm X_{i,j}$ to give $\pm A_\diamond := \pm Z_{i-1,j} Z_{i+1,j} Z_{i,j-1} Z_{i,j+1}$. The $A_\diamond$ stabilizers exist on the blue sublattice, which is in fact another square lattice, rotated by 45$^\circ$. Therefore, the resulting state on the blue sublattice is an eigenstate of the Xu\nobreakdash-Moore Hamiltonian~\cite{Xu2004},
\begin{equation}
  H = - \sum A_\diamond.
\end{equation}
This self-dual model arose as a proposed description of superconducting arrays~\cite{Xu2004, Xu2005}, but also is often connected to the transverse-field toric code~\cite{Vidal2009}. The Xu-Moore Hamiltonian is equivalent to the quantum compass model~\cite{Nussinov2005}, which is known to exhibit topological protection of qubits~\cite{Doucot2005, Dorier2005, Fernandez-Lorenzo2016}: specifically, the topological phase is protected against the introduction of a transverse field.

In the stochastic case of $p_u=1$ and varying $p_m$, we show that the circuit depth to recover a Xu-Moore state scales logarithmic with system size [see Fig.~\ref{fig:2d-square-results}(a)]. The $A_\diamond$ stabilizers, when produced by measurements in the odd layers, are stable to further evolution. The mean time $\bar\tau$ should therefore be bounded from above by the time needed to measure each of the $L^2/2$ sites at least once, which is indeed logarithmic. 

However, near $p_m \to 1$, we find significant deviations from purely logarithmic behavior, which can be understood as follows. First, note that $A_\diamond$ stabilizers can be joined together along a diagonal or in a cone shape, mimicking the global symmetries. One missing $A_\diamond$ can naturally arise given all other stabilizers are present on a diagonal or in a cone. When $p_m \to 1$, many $A_\diamond$ stabilizers will be present after a few time steps due to random measurements, and even more can arise due to the aforementioned process. This leads to the pinning of the mean time $\bar\tau$ to values close to $(2n+1), n\in\mathbb{Z}$ (where the factor of 2 is again due to the even-odd layer effects, similar to Sec.~\ref{sec:cat-pu1}). We examine this understanding by simulating a coin-toss experiment, where we toss $L^2/2$ coins, fix the heads results, check if any tails should be flipped due to a diagonal or a cone, and repeat until all coins are heads. The results are plotted as solid lines in Fig.~\ref{fig:2d-square-results}(a) and match the numerical data. Again, we expect that when both $p_u$ and $p_m$ are below unity, the time needed to achieve the Xu-Moore state diverges exponentially. However, we also speculate that the halting protocol should reduce this time to an efficiently simulable time scale.

\subsection{Measurement-induced entanglement transition in two dimensions}

Having extended our stochastic protocol results to two dimensions, where we discovered the ability to prepare states with topological order, we now turn our attention to the possibility of finding a measurement-induced transition akin to the one-dimensional case. To explore this, we employ the non-commuting gates described by Eq.~(\ref{eq:unitary-miet}) and utilize the Lieb lattice depicted in Fig.~\ref{fig:2d-lattices}(a). Our approach involves applying the gates in four layers, following the pattern: apply the unitary gates $U[(i,j),(i,j+1)]$, followed by $U[(i,j),(i,j-1)]$, $U[(i,j),(i+1,j)]$, and finally $U[(i,j),(i-1,j)]$. In this context, $(i,j)$ represents a red site, and $U[a,b]$ denotes a unitary operation applied on sites with indices $a$ and $b$. Subsequently, measurements are exclusively performed on the red sites. For the calculation of the tripartite mutual information $I_3$, we define the regions to be strips of size $L \times (L/4)$.

\begin{figure}[t]
  \centering
  \includegraphics[width = 0.91\columnwidth]{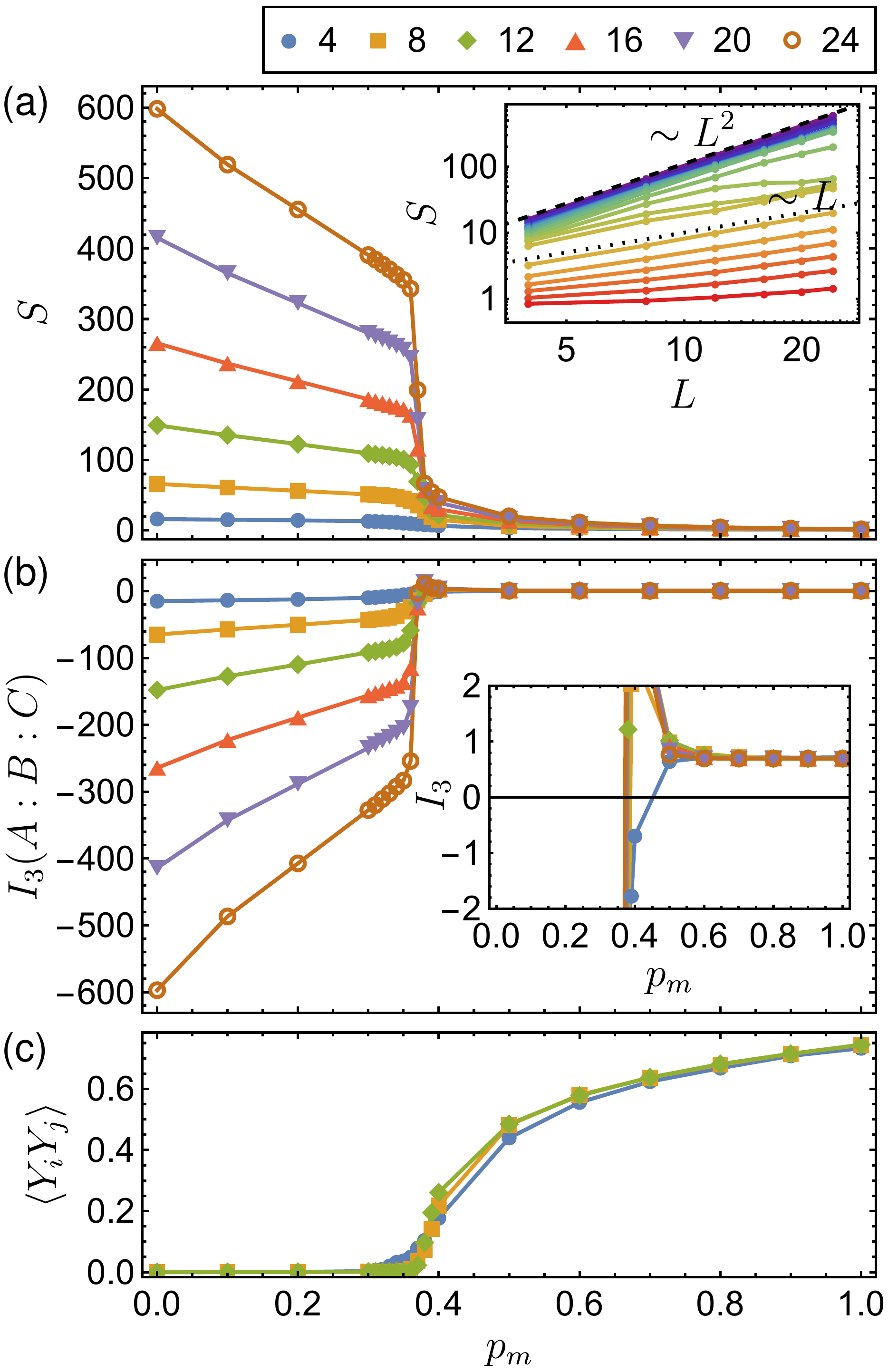}
  \caption{Measurement-induced transition present on a two-dimensional Lieb lattice, with the unitary evolution including both ZZ and XX terms. (a)~Half-plane entanglement entropy as a function of the measurement probability. The inset shows a dependence on the linear system dimension $L$, colors corresponding to different $p_m$ (small values in blue, large values in red). Dashed line shows a slope of $\sim L^2$, while dotted line shows a slope of $\sim L$. (b)~Tripartite mutual information $I_3(A:B:C)$ with a clear signature of a phase transition. Each region is a rectangle of size $L/4\times L$. The inset shows a zoomed-in version of the plot, where $I_3$ is near $\log(2)$ in the area-law phase. (c)~Expectation value of local stabilizers of the 2D cat state $|\langle Y_i Y_j \rangle|$. In all plots, $p_u = 0.9$.
  \label{fig:2d-square-miet}}
\end{figure}

As presented in Fig.~\ref{fig:2d-square-miet}, the results indicate the presence of an entanglement transition between a volume law phase with half-plane entropy $S\sim L^2$ and an area law phase with $S\sim L$. In the area law phase, the quantum state closely resembles a cat state oriented along the $Y$ direction, with $I_3$ approaching $\log(2)$, and with an increased likelihood of generating the $Y_i Y_j$ stabilizers associated with the cat state (where $i$ and $j$ are unmeasured sites). In contrast, the volume law is featureless, with the $Y_i Y_j$ stabilizers occurring with a vanishing probability in the infinite volume limit. Hence, the results generalize the effects observed in one dimension, revealing an underlying universality of our approach. Indeed, we believe that a similar protocol can be leveraged to design entanglement transitions in many different lattices, where the area law is constructed to resemble a long-range entangled state.

Summarizing this section, we find that in two dimensions, our stochastic protocol involving unitaries and measurements is endowed with state preparation time scales consistent with higher-dimensional generalization of one dimension. In this case, the target states are an error-correcting toric code state and a Xu-Moore state. We specifically show the existence of a logarithmic time scale, where local symmetries of desired states are recovered, and exponential long times to recover global symmetries. Owing to the higher dimensionality of the problem, the number of global symmetries is larger (extensive in the system size), and hence, the recovery of the target state is harder. However, the halting protocol can be used to produce the target states in a viable amount of time. Furthermore, we find that by using non-commuting gates one can force a measurement-induced entanglement transition in a two-dimensional lattice, characterized by an area law reminiscent of a cat state. Our 2D results suggest that our findings for the stochastic protocol in one dimension have a strong degree of universality and can be used to design similar protocols on other geometries.

\section{Discussion}
\label{sec:discussion}

In this work, we explored the effects of imperfections and stochasticity in state preparation protocol\changes{s} for long-range entangled states consisting of unitary gates and measurements in one and two dimensions. \changes{We propose a general mechanism that describes the dynamical stability of these protocols under a wide range of perturbations.}  Our main result is the emergence of two timescales for achieving the desired state: a time scale logarithmically diverging in system size, which we relate to the presence of local symmetries in the system, and an exponentially diverging time required to recover the global symmetries of the desired state. \changes{The relationship between timescales and symmetries provides a fundamental constraint on the stability of the protocols.} We investigate methods for speeding up the protocol, such as a local decoder and a halting protocol and find \changes{the latter to be significantly more efficient}, which tackles the exponential growth of time scales. Surprisingly, our stochastic protocol is relatively stable in the presence of timing imperfections, where our state may no longer be a stabilizer state. Moreover, the insertion of an additional interaction term in the evolution leads to the emergence of a measurement-induced entanglement transition, where for a high frequency of measurements we find that the steady state behaves similarly to a cat state. Through our results, we are able to probe the thresholds on experimental errors and give insights for efficient error correction. \changes{The similarity of our results across one and two dimensions provides a solid basis for the generality of the two timescales observed in this study, while the entanglement transition serves as a universal measure of the robustness of the state preparation protocol.}

An extension of our work could involve the inclusion of error correction of logical qubits in stochastic circuits. In particular, whether efficient correction can be performed with an efficient scaling of the overhead, specifically so that the additional resources scale sub-extensively~\cite{Bonet2018, livingston2022}. Novel methods of quantum error correction using additional flag qubits~\cite{Chao2018} could provide an alternate route with a scalable complexity of the protocol. At a broader level, classification of the complexity of the preparation protocols could be more rigorously defined, and in particular the relationship of the protocol to no-go theorems for resource purification~\cite{Fang2020}\changes{. Another interesting avenue for future explorations involves incorporating feedback into the protocol with non-commuting gates. Feedback in systems exhibiting entanglement transitions may lead to an absorbing-state transition~\cite{Buchhold2022, ODea2024, Sierant2023, Iadecola2023, Friedman2023, Ravindranath2023}, which could be leveraged to stabilize our target long-range entangled state.}

Investigations of the stochastic protocol in higher dimensions open new exciting directions. Preparation of multi-particle anyon states in the toric code can be a first step to a realization of anyon manipulation and direct investigations of anyonic braiding processes, especially since our protocol prepares a well-defined string operator. This has direct implications in the highly active field of topological quantum computing~\cite{Andersen2022, Iqbal2023feedforward, Bartolomei2020, Nakamura2020, Iqbal2023}. Secondly, one can pose a question of the quantum fault-tolerance threshold for anyons, a natural problem to be investigated within the context of stochastic circuits with measurements. Furthermore, it would be interesting to investigate the existence of a local decoder for the 2D stochastic surface code protocols, a topic that has received some recent attention~\cite{Breuckmann2016, Breuckmann2018, Kubica2019, Kubica2023}. The development of such a decoder could help to stabilize anyonic particles under random processes and make them viable platforms for investigating anyon statistics in experimental setups.

Generalizing the projective measurement protocols for state preparation to weak measurements can be advantageous for experiments~\cite{murch2013, hatridge2013, Roch2014}. Weak measurements to steer the state close to the manifold of the target state could provide scalable protocols which provide a shorter preparation time given a fidelity for the long-range entangled state~\cite{Zhu2022, Lee2022, Lee2023}. These factors are relevant for superconducting circuits with weak measurements and provide a realizable model for stochastic measurements. Theoretical implications of weak measurements may allow the development of protocols with unitary evolution  not given by frustration-free elements~\cite{Cubitt2023} relevant to classify the complexity of dynamics of open quantum systems.

\begin{acknowledgments}
M.S.\@ and A.P.\@ were funded by the European Research Council (ERC) under the European Union's Horizon 2020 research and innovation programme (Grant Agreement No.\@ 853368). The authors acknowledge the use of the UCL Myriad High Performance Computing Facility (Myriad@UCL), and associated support services in the completion of this work.

All relevant data present in this publication can be accessed at~\cite{researchdata}.
\end{acknowledgments}

\appendix

\section{Logarithmic times in the stochastic circuit}
\label{sec:appendix-log-time}

Here we provide a precise calculation of the mean time to reach a cat state $\bar\tau$, which can be performed by treating each application of neighboring unitaries separately as a random variable. Firstly, let us analyze the case of $p_m=1$, illustrated in Fig.~\ref{fig:circuit-pm1}. The mean time to the first occurrence of two neighboring unitaries is described by the geometric distribution with probability mass function (PMF) $f(t)$ and cumulative distribution function (CDF) $F(t)$,
\begin{equation}
  f (t) = (1 - p_u^2)^t p_u^2, \qquad F (t) = 1 - (1 - p_u^2)^{t + 1} .
\end{equation}
The mean time to get a cat state is then described by the situation when we choose the 2nd largest time out of $L / 2$ trials (as the last $Z_i Z_{i+2}$ stabilizer follows from the global symmetry). The PMF of this new distribution $f_{\text{os}}^{(p_u^2)}(t)$ is given by a formula for the $(L/2 - 1)$-th order statistic,
\begin{align}
  & f_{\text{os}}^{(p_u^2)}(t)\nonumber\\
  &=  \sum_{j = 0}^{1} \binom{L / 2}{j} \left( (1 - F (t))^j (F (t))^{L / 2 - j}\vphantom{\frac{}{}} \right. \nonumber\\
  & \quad - \left. \vphantom{\frac{}{}} (1 - F (t) + f (t))^j (F (t) - f (t))^{L / 2 - j} \right) \label{eq:order-statistic} \\
  & =  (1 - (1 - p_u^2)^{t + 1})^{L/2 - 1} \left( 1 + \left( \tfrac{L}{2} - 1 \right) (1 - p_u^2)^{t + 1} \right) \nonumber\\
  &  \quad - (1 - (1 - p_u^2)^t)^{L/2 - 1} \left( 1 + \left( \tfrac{L}{2} - 1 \right) (1 - p_u^2)^t \right) . 
\end{align}
The mean time $\bar\tau$ is therefore
\begin{align}
  \bar\tau&=\sum_{t = 0}^{\infty} t f_{\text{os}}^{(p_u^2)} (t) + 1 \\
  & \approx   \frac{1 - B_{(p_u^2)} (1 + \frac{L}{2}, 0) - H (\frac{L}{2})}{\log (1 - p_u^2)} - \frac{p_u^L}{2} + \frac{1}{2}\quad\label{eq:exact-mean-time-pm1}\\
  & \xrightarrow[L \to \infty]{}  \frac{\log (2 / L)}{\log (1 - p_u^2)} - \frac{\gamma - 1}{\log (1 - p_u^2)} + \frac{1}{2},
\end{align}
where $B_x (a, b) = \int_0^x u^{a - 1} (1 - u)^{b - 1} d u \xrightarrow[a \to \infty]{} \frac{x^a}{a} \sum_{k = 0}^{\infty} \frac{1}{a^k} \frac{d^k}{d x^k} ((1 - x e^{- w})^{b - 1})_{w = 0} \approx 0$ is the incomplete Beta function and $H (n) = \sum_{i = 1}^n \frac{1}{i} \xrightarrow[n \to \infty]{} \gamma + \log (n)$ is the Harmonic number. The first approximation is done by rewriting a sum as an integral, while the second approximation is a series expansion at $L \to \infty$. Note that this expansion has the same leading term as the naive calculation from Eq.~(\ref{eq:naive-mean-time}).

\begin{figure}[tb]
  \centering
  \includegraphics[width = 0.99\columnwidth]{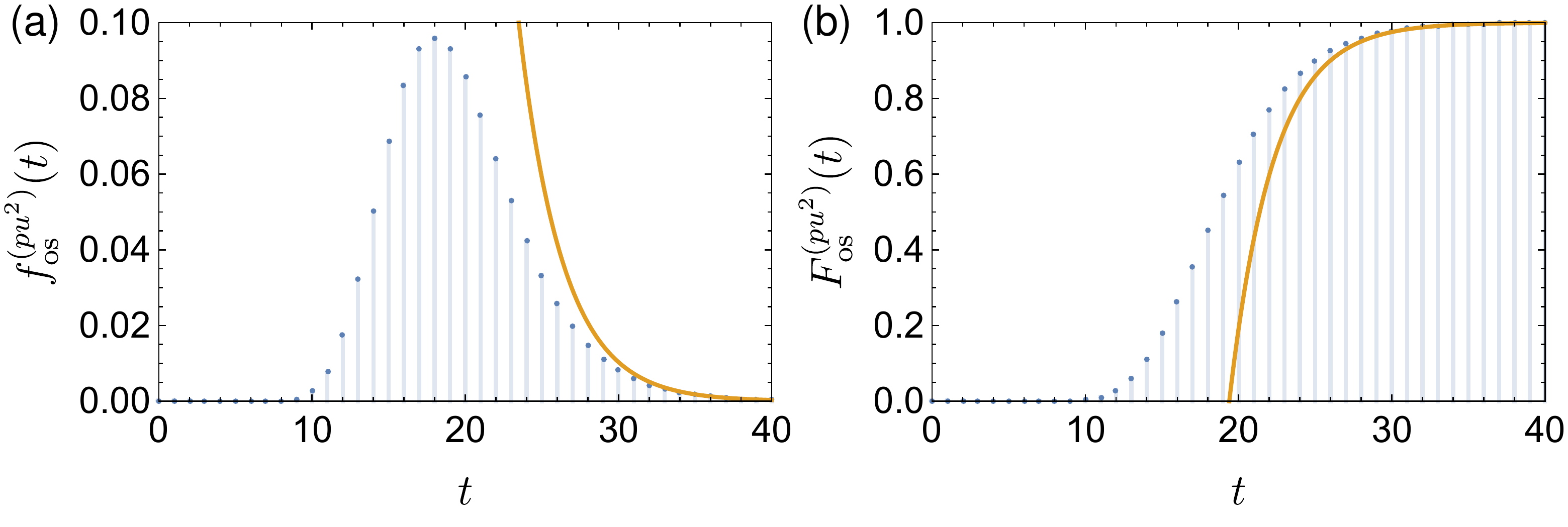}
  
  \caption{(a)~Probability mass function $f_{\text{os}}^{(p_u^2)}(t)$ and (b)~cumulative distribution function $f_{\text{os}}^{(p_u^2)}(t)$ of time to reach a cat state (example for $L = 100, p_m = 1, p_u = 0.4$). Orange lines show the asymptotic exponential behavior. 
  \label{fig:time-distribution}}
\end{figure}

An example of the PMF of the time distribution is shown in Fig.~\ref{fig:time-distribution}. The distribution decays exponentially in time according to:
\begin{align}
  f_{\text{os}}^{(p_u^2)} (t) &\xrightarrow[t \to \infty]{} \frac{1}{8} (L - 2) L (2 - p_u^2) p_u^2 (1 - p_u^2)^{2 t} \\ &\sim \exp (- a t).
\end{align}
Similarly, the CDF is exponentially approaching 1 according to:
\begin{align}
  F_{\text{os}}^{(p_u^2)} (t) &\xrightarrow[t \to \infty]{} 1 - \frac{1}{8} (L - 2) L (1 - p_u^2)^{2 (t + 1)} \\ &\sim 1 - \exp (- b t),
\end{align}
which can be interpreted as follows: in order to achieve the fidelity of $\phi$, one needs to reach the times of
\begin{equation}
  \tau(\phi, p_u^2) \sim \frac{1}{2}  \frac{\log \left( \frac{8 (1 - \phi)}{L (L - 2)} \right)}{\log (1 - p_u^2)} - 1.
\end{equation}

The case of $p_u = 1$ is similar to that of $p_m=1$, however, one needs to account for the even-odd layer difference in producing stable $Z_{i - 1} Z_{i + 1}$ stabilizers. This leads to the mean time $\bar\tau$ of
\begin{align}
  \bar\tau & =  2 \sum_{t = 0}^{\infty} t f_{\text{os}}^{(p_m)} (t) + 1 \\
  & \approx  2\,\frac{1 - B_{p_m} (1 + \frac{L}{2}, 0) - H (\frac{L}{2})}{\log (1 - p_m)} - p_m^{L / 2} \label{eq:exact-mean-time-pu1}\\
  & \xrightarrow[L \to \infty]{}  2 \left( \frac{\log (2 / L)}{\log (1 - p_m)} - \frac{\gamma - 1}{\log (1 - p_m)} \right).
\end{align}

\begin{table}[t]
    \footnotesize
    \centering
    \setlength\tabcolsep{3pt}
  \begin{tabular}{cccc}
    \hline \hline
    Circuit & Probability & Stabilizer change & Outcome\\
    \hline
    \includegraphics[valign=m,width=1.7cm]{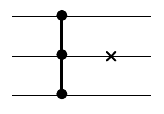} &
    $p_u^2 p_m$ &
    ${\arraycolsep=1.4pt
    \begin{array}{rl}
      X_2 &\rightarrow X_2\\
      Y_2 Z_3 &\rightarrow X_2 \phantom{Z_3}
    \end{array}}$ & Rand.
    \\
    \hline
    \includegraphics[valign=m,width=1.7cm]{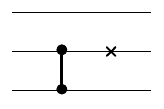} &
    $p_u (1 {-} p_u) p_m$ &
    ${\arraycolsep=1.4pt
    \begin{array}{rl}
      X_2 &\rightarrow X_2\\
      Y_2 Z_3 &\rightarrow X_2 \phantom{Z_3}
    \end{array}}$ & Rand.
    \\
    \hline
    \includegraphics[valign=m,width=1.7cm]{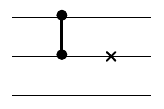} &
    $p_u (1 {-} p_u) p_m$ &
    ${\arraycolsep=1.4pt
    \begin{array}{rl}
      X_2 &\rightarrow X_2\\
      Y_2 Z_3 &\rightarrow X_2 \phantom{Z_3}
    \end{array}}$ & Rand.
    \\
    \hline
    \includegraphics[valign=m,width=1.7cm]{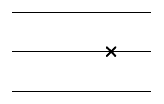} &
    $(1 {-} p_u)^2 p_m$ &
    ${\arraycolsep=1.4pt
    \begin{array}{rl}
      X_2 &\rightarrow X_2\\
      Y_2 Z_3 &\rightarrow X_2 \phantom{Z_3}
    \end{array}}$ & Determ.
    \\
    \hline
    \includegraphics[valign=m,width=1.7cm]{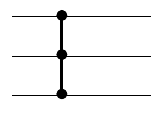} &
    $p_u^2 (1 {-} p_m)$ &
    ${\arraycolsep=1.4pt
    \begin{array}{rl}
      X_2 &\rightarrow X_2\\
      Y_2 Z_3 &\rightarrow Y_2 Z_3
    \end{array}}$ & Rand.
    \\
    \hline
    \includegraphics[valign=m,width=1.7cm]{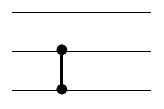} &
    $p_u (1 {-} p_u) (1 {-} p_m)$ &
    ${\arraycolsep=1.4pt
    \begin{array}{rl}
      X_2 &\rightarrow Y_2 Z_3\\
      Y_2 Z_3 &\rightarrow X_2
    \end{array}}$ & Rand.
    \\
    \hline
    \includegraphics[valign=m,width=1.7cm]{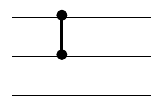} &
    $p_u (1 {-} p_u) (1 {-} p_m)$ &
    ${\arraycolsep=1.4pt
    \begin{array}{rl}
      X_2 &\rightarrow Y_2 Z_3\\
      Y_2 Z_3 &\rightarrow X_2
    \end{array}}$ & Rand.
    \\
    \hline
    \includegraphics[valign=m,width=1.7cm]{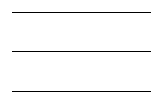} &
    $(1 {-} p_u)^2 (1 {-} p_m)$ &
    ${\arraycolsep=1.4pt
    \begin{array}{rl}
      X_2 &\rightarrow X_2\\
      Y_2 Z_3 &\rightarrow Y_2 Z_3
    \end{array}}$ & Rand.
    \\
    \hline \hline
  \end{tabular}
  \caption{All possible 3-qubit one-time-step circuits and the corresponding probabilities. The third column shows how the circuit transforms $X_2$ and $Y_2 Z_3$ stabilizers [see the discussion of Eq.~(\ref{eq:p_X})]. The fourth column shows whether the measurement outcome is random or deterministic, if one starts from the separable $|-\rangle^{\otimes L}$ state.
  \label{tab:local-circuits}}
\end{table}

In the 2D Lieb lattice, a similar process can be used to estimate the mean time when $p_u=1$. $(L^2-1)$ stabilizers need to be fixed, thus resulting in
\begin{align}
  \bar\tau & =  2 \sum_{t = 0}^{\infty} t f_{\text{os}}^{(p_m)} (t) + 1 \\
  & \approx  2\,\frac{1 - B_{p_m} (1 + L^2, 0) - H (L^2)}{\log (1 - p_m)} - p_m^{L^2} \label{eq:lieb-mean-time-pu1} \\
  & \xrightarrow[L \to \infty]{}  2 \left( \frac{\log (1/L^2)}{\log (1 - p_m)} - \frac{\gamma - 1}{\log (1 - p_m)} \right).
\end{align}

\section{Exponential times in the stochastic circuit}
\label{sec:appendix-exp-time}

In this section, we give a detailed derivation of the mean time to reach a cat state $\bar \tau$ for the case when both $p_u$ and $p_m$ are below 1. This time consists of two parts: time $\bar\tau_\ZZ$ to achieve all local $Z_i Z_{i+2}$ stabilizers, and time $\bar\tau_{\mathbb{Z}_2}$ to recover the global $\mathbb{Z}_2$ symmetry of $\prod_{i \text{ odd}} X_i$.

Naively, one could estimate the mean time needed to recover all the $Z_i Z_{i + 2}$ stabilizers using a similar procedure as for $p_m=1$ or $p_u=1$, by assuming that the necessary stabilizers are produced any time two neighboring unitaries are followed by measurements. This would then lead to
\begin{equation}
  \bar\tau_{\ZZ}=\sum_{t = 0}^{\infty} t f_{\text{os}}^{(p_u^2 p_m)}(t)+1 \sim \frac{\log (2 / L)}{\log (1 - p_u^2 p_m)}.
\end{equation}
However, the scenarios leading to a stable $Z_i Z_{i + 2}$ symmetry are more complicated and the naive estimation fails at small times. We instead should consider all possible 3-site one-time-step circuits (see Table~\ref{tab:local-circuits}) and their effects on the local stabilizers. We find six 3-qubit states which are possible during the evolution, and here we list their stabilizer generators:
\begin{align}
  (1) & \qquad \langle X_1, X_2, X_3 \rangle, \\
  (2) & \qquad \langle X_1, Z_2 Y_3, X_2 X_3 \rangle, \\
  (3) & \qquad \langle X_3, Z_2 Y_1, X_2 X_1 \rangle, \\
  (4) & \qquad \langle Y_1 Z_2, Z_2 Y_3, Z_1 X_2 Z_3 \rangle, \\
  (5) & \qquad \langle Z_1 Z_3, X_2, X_1 X_2 X_3 \rangle, \label{eq:state5}\\
  (6) & \qquad \langle Z_1 Z_3, Y_2 Z_3, X_1 X_2 X_3 \rangle. \label{eq:state6}
\end{align}
Transitions between the different states can be summarized in the graph in Fig.~\ref{fig:transfer-graph}, where each edge has a weight corresponding to the probability of the transition. This weighted graph has the following adjacency matrix $\mathcal{A}$,
\begin{widetext}
\begin{equation}
\mathcal{A} = {\footnotesize\medmuskip=1mu
\begin{pmatrix}
 2 p_m p_u q_u+q_u^2 & p_m p_u^2+p_m q_u^2+p_u q_u & p_m p_u^2+p_m q_u^2+p_u q_u & 2 p_m q_u p_u+p_u^2 & 0 & 0 \\
 q_m q_u p_u & q_m q_u^2 & q_m p_u^2 & q_m q_u p_u & 0 & 0 \\
 q_m q_u p_u & q_m p_u^2 & q_m q_u^2 & q_m q_u p_u & 0 & 0 \\
 q_m p_u^2 & q_m q_u p_u & q_m q_u p_u & q_m q_u^2 & 0 & 0 \\
 p_m p_u^2 & p_m q_u p_u & p_m q_u p_u & p_m q_u^2 & 2 p_m p_u q_u+p_u^2+q_u^2 & p_m p_u^2+p_m q_u^2+2 p_u q_u \\
 0 & 0 & 0 & 0 & 2 q_m q_u p_u & q_m p_u^2+q_m q_u^2
\end{pmatrix}},
\end{equation}
\end{widetext}
where $q_u = 1-p_u$ and $q_m=1-p_m$. Element $\mathcal{A}_{i,j}$ corresponds to the probability of transition from state ($j$) to state ($i$).

Note that states 5 and 6 from Eqs. (\ref{eq:state5}) and (\ref{eq:state6}) are stable at long times and both have a $Z_i Z_{i+2}$ stabilizer. As a function of time $t$, the CDF of obtaining a $Z_i Z_{i+2}$ stabilizer is therefore:
\begin{equation}
  F_\ZZ (t) = (0\ \ 0\ \ 0\ \ 0\ \ 1\ \ 1)\,\mathcal{A}^t\, (1\ \ 0\ \ 0\ \ 0\ \ 0\ \ 0)^T, \label{eq:CDF_ZZ}
\end{equation}
which takes into account that one starts from state 1 and ends at state 5 or 6. This then can be used in the formula for the order statistic from Eq.~(\ref{eq:order-statistic}), and finally, to obtain the mean time $\bar\tau_\ZZ$.

After reaching $Z_{i} Z_{i + 2}$ stabilizers on every pair of odd sites, we find that the $\mathbb{Z}_2$ symmetry of $\prod_{i \text{ odd}} X_i$ needed for the cat state may still be absent (in fact, for larger systems it is almost always absent after $\bar\tau_\ZZ$). We note that only two types of states are possible: either state 5 with stabilizer $X_i$, or state 6 with stabilizer $Y_i Z_{i + 1}$ for every even $i$. To recover the $\mathbb{Z}_2$ symmetry of $\prod_{i \text{ odd}} X_i$ needed for the cat state, one can combine the global $\prod_{i = 0}^{L - 1} X_i$ symmetry with local $X_i$ stabilizers (state 5), assuming they are present on every possible even site. The mean time to this situation $\bar\tau_{\mathbb{Z}_2}$ can be calculated as follows. Using the transitions corresponding to local circuits from Table~\ref{tab:local-circuits}, one can write the following equation involving the probability $p_X$ of obtaining one local $X_i$ stabilizer (i.e.\@ obtaining state 5):
\begin{align}
  p_X &= p_X ((1 - p_u)^2 (1 - p_m) + p_u^2 (1 - p_m))\nonumber \\
  & \quad + 2 (1 - p_X) p_u (1 - p_u) (1 - p_m)\nonumber\\
  & \quad + p_m,
  \label{eq:p_X}
\end{align}
where the first term corresponds to obtaining $X_i$ while starting from $X_i$ and not applying a measurement, the second term corresponds to obtaining $X_i$ while starting from $Y_i Z_{i+1}$ and not applying a measurement, and the last term corresponds to a circuit with a measurement that leads to an $X_i$ stabilizer. Therefore,
\begin{equation}
  p_X = \frac{p_m + 2 (p_m - 1) p_u (p_u - 1)}{p_m + 4(p_m - 1)p_u(p_u-1)}.
\end{equation}
The time $\bar\tau_{\mathbb{Z}_2}$ is distributed according to geometric
distribution, which has a mean of
\begin{align}
  \bar\tau_{\mathbb{Z}_2} &\sim 1 / p_X^{L / 2} = \left( \frac{p_m + 4 (p_m - 1) p_u (p_u - 1)}{p_m + 2 (p_m - 1) p_u (p_u - 1) } \right)^{L / 2}.
\end{align}

\section{Stability under transverse field}
\label{sec:appendix-stability-transverse-field}

\begin{figure}[tb]
  \centering
  \includegraphics[width = 0.99\columnwidth]{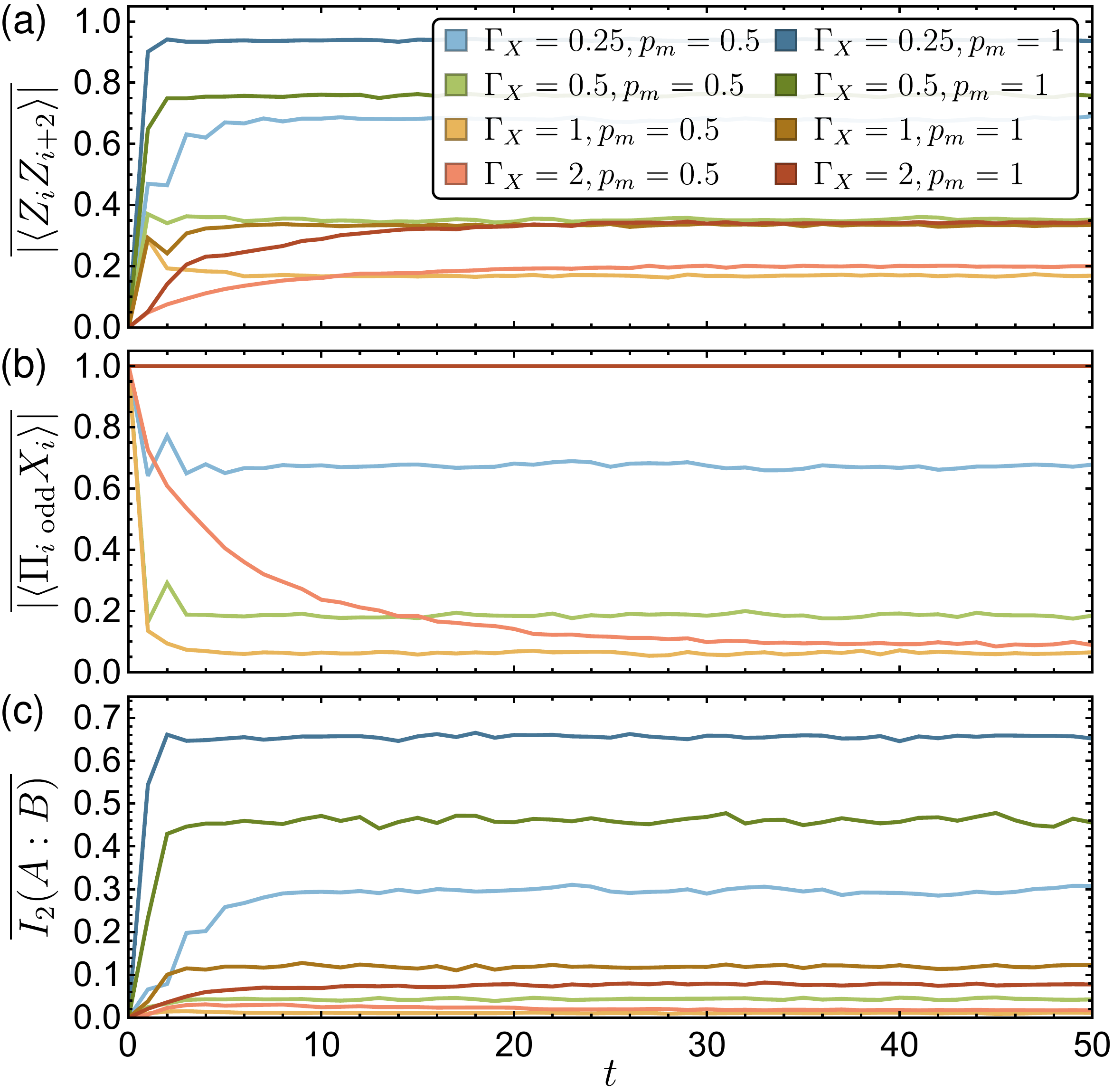}
  \caption{Protocol with an additional transverse field term: averaged expectation value of (a)~stabilizers $\langle Z_i Z_{i+2} \rangle$, (b)~the global symmetry $\langle \prod_{i\text{ odd}} X_i \rangle$, and the mutual information between two antipodal unmeasured sites A and B, $I_2(A:B)$. Legend in (a) applies to (b) and (c). The system size is $L=14$.
  \label{fig:transverse-field-term}
  }
\end{figure}

We investigate a protocol where the Hamiltonian in Eq.~(\ref{eq:unitary}) acquires a transverse field term. The new Hamiltonian is given by $H = \sum_i Z_i Z_{i + 1} + \Gamma_X \sum_i X_i$. The unitary evolution can no longer be split into two-site unitaries, and instead, to efficiently implement $\exp(-i \Delta t H) |\psi\rangle$, we use a dense method \changes{(where the state is represented as a vector of $2^L$ complex coefficients)} based on exponential integrators~\cite{AlMohy2011}.

We find (see Fig.~\ref{fig:transverse-field-term}) that the transverse field generically leads to a steady state that is not a cat state. When $p_m=1$, we note that the evolution preserves the global $\prod_{i \text{ odd}} X_i$ symmetry, but there is no stable $Z_i Z_{i + 2}$ symmetry.

\section{Analytical results for the stability under time imperfections}
\label{sec:appendix-time-imperfections-proof}

In this section, we aim to prove that when using unitary gates with time imperfections, the expectation value $| \langle \psi | Z_{i} Z_{i+2} | \psi \rangle|$, averaged over applications of unitaries, measurements, and measurement outcomes, does not decay and can only grow or stay constant. Without the loss of generality, we focus on $i=1$, i.e.\@ expectation value $| \langle \psi | Z_1 Z_3 | \psi \rangle|$. We prove this for a one-time-step circuit on a 3-qubit cluster, and therefore, by extension, for the entire evolution.

Let us consider a representative contribution to the statistical average, where both the unitaries and the measurement are present in the system,
\begin{align}
  p_m & p_u^2 \sum_{s \in \{ +, - \}} p_s  \frac{| \langle \psi' | Z_1 Z_3 | \psi'
  \rangle |}{| \langle \psi' | \psi' \rangle |} \nonumber\\ 
  &= p_m p_u^2 \ \smashoperator{\sum_{s \in \{ +, - \}}} \  p_s  \frac{| \langle \psi | U^{\dag}_{12} U^{\dag}_{23} P^s_2 Z_1 Z_3
  P^s_2 U_{12} U_{23} | \psi \rangle |}{| \langle \psi | U^{\dag}_{12}
  U^{\dag}_{23} P^s_2 U_{12} U_{23} | \psi \rangle |},
\end{align}
where $| \psi \rangle$ is the initial wave function, $| \psi' \rangle$ is the wave function after one time step, and the only sum left is over measurement outcomes $s$. $P_i^s = |s\rangle\langle s|$ is the projector corresponding to the outcome $s$. Noticing that the probability of the measurement outcome is $p_s = | \langle \psi | U^{\dag}_{12} U^{\dag}_{23} P^s_2 U_{12} U_{23} | \psi \rangle |$, one can use the triangle inequality to show that
\begin{align}
  p_m & p_u^2 \sum_{s \in \{ +, - \}} p_s  \frac{| \langle \psi' | Z_1 Z_3 |
  \psi' \rangle |}{| \langle \psi' | \psi' \rangle |} \nonumber \\ 
  & \geqslant p_m p_u^2
  \left| \langle \psi | U^{\dag}_{12} U^{\dag}_{23} \left( \quad\smashoperator{\sum_{s \in \{ +, -
  \}}} P^s_2 \right) Z_1 Z_3 U_{12} U_{23} | \psi \rangle \right| \\
  & = p_m p_u^2 | \langle \psi | Z_1 Z_3 | \psi \rangle | . 
\end{align}
Averaging over applications of measurements and unitaries, we conclude that
\begin{equation}
  \mathbb{E}  | \langle \psi' | Z_1 Z_3 | \psi' \rangle | \geqslant | \langle
  \psi | Z_1 Z_3 | \psi \rangle |,
\end{equation}
where $\mathbb{E} (x)$ designates an average over all possible trajectories.

\bibliographystyle{quantum}
\bibliography{refs}

\end{document}